\documentclass[twocolumn,epjc3]{svjour3}  
\pdfoutput=1
\RequirePackage{fix-cm}
\RequirePackage{amsmath}
\RequirePackage{amssymb}
\RequirePackage{graphicx}
\RequirePackage{mathptmx}      

\RequirePackage{latexsym}
\RequirePackage[numbers,sort&compress]{natbib}
\RequirePackage[colorlinks,citecolor=blue,urlcolor=blue,linkcolor=blue]{hyperref}

\journalname{Eur. Phys. J. C}

\usepackage{xspace}

\newcommand{\Sherpa}{\protect\textsc{Sherpa}\xspace}
\newcommand{\Amegic}{\protect\textsc{Amegic}\xspace}
\newcommand{\Recola}{\protect\textsc{Recola}\xspace}
\newcommand{\Collier}{\protect\textsc{Collier}\xspace}
\newcommand{\LHAPDF}{\protect\textsc{Lhapdf}\xspace}
\newcommand{\Rivet}{\protect\textsc{Rivet}\xspace}
\newcommand{\Fastjet}{\protect\textsc{FastJet}\xspace}

\newcommand{\order}{\ensuremath{\mathcal{O}}}
\newcommand{\done}{\ensuremath{\mathrm{d}}}
\newcommand{\alphaS}{\ensuremath{\alpha_s}}
\newcommand{\Gmu}{\ensuremath{G_\mu}}
\newcommand{\shortequal}{\ensuremath{\!\!\!\!\!\!=\!\!\!\!\!\!}}
\newcommand{\pT}{\ensuremath{p_\mathrm{T}}}
\newcommand{\GeV}{\ensuremath{\text{GeV}}}
\newcommand{\TeV}{\ensuremath{\text{TeV}}}
\newcommand{\HThalf}{\ensuremath{\tfrac{1}{2}\,\HT}}
\newcommand{\HT}{\ensuremath{\hat{H}_\mathrm{T}}}
\newcommand{\HTtwo}{\ensuremath{H_\mathrm{T}^{(2)}}}
\newcommand{\Rtt}{\ensuremath{R_{32}}}

\newcommand{\LO}{\ensuremath{\text{LO}}}
\newcommand{\QCD}{\ensuremath{\text{QCD}}}
\newcommand{\EW}{\ensuremath{\text{EW}}}
\newcommand{\LOzero}{\ensuremath{\LO_{0}}}
\newcommand{\LOone}{\ensuremath{\LO_{1}}}
\newcommand{\LOtwo}{\ensuremath{\LO_{2}}}
\newcommand{\LOthree}{\ensuremath{\LO_{3}}}

\newcommand{\NLO}{\ensuremath{\text{NLO}}}
\newcommand{\DNLO}{\ensuremath{\Delta\NLO}}
\newcommand{\DNLOzero}{\ensuremath{\Delta\NLO_{0}}}
\newcommand{\DNLOone}{\ensuremath{\Delta\NLO_{1}}}
\newcommand{\DNLOtwo}{\ensuremath{\Delta\NLO_{2}}}
\newcommand{\DNLOthree}{\ensuremath{\Delta\NLO_{3}}}
\newcommand{\DNLOfour}{\ensuremath{\Delta\NLO_{4}}}

\newcommand{\QCDpEW}{\ensuremath{\text{QCD}+\text{EW}}}
\newcommand{\QCDtEW}{\ensuremath{\text{QCD}\times\text{EW}}}

\newcommand{\hl}{\ensuremath{\vphantom{\int_A^B}}}
\newcommand{\Hl}{\ensuremath{\vphantom{\int\limits_A^B}}}

\newlength{\unitcharwidth}
\hypersetup{
  pdfauthor={Max Reyer, Marek Schoenherr, Steffen Schumann},
  pdftitle={Full NLO corrections to 3-jet production and R32 at the LHC}
}

\begin{document}

\title{Full NLO corrections to 3-jet production and $\mathbf{\Rtt}$ at the LHC}

\author{Max Reyer\thanksref{e1,addr1,addr3}
        \and
        Marek Sch\"onherr \thanksref{e2,addr2,addr2a}
        \and
        Steffen Schumann \thanksref{e3,addr3}
	\begin{picture}(0,0)
	  \put(-66,75){FR-PHENO-2018-016, CERN-TH-2018-275, IPPP/19/9, MCNET-19-03}
	\end{picture}
}

\thankstext{e1}{e-mail: max.reyer@physik.uni-freiburg.de}
\thankstext{e2}{e-mail: marek.schoenherr@cern.ch}
\thankstext{e3}{e-mail: steffen.schumann@phys.uni-goettingen.de}


\institute{Albert-Ludwigs-Universit\"at Freiburg, Physikalisches Institut, Hermann-Herder-Stra\ss e 3, 79104 Freiburg, Germany \label{addr1}
           \and
           Theoretical Physics Department, CERN, 1211 Geneva 23, Switzerland\label{addr2}
           \and
           Institute for Particle Physics Phenomenology, University of Durham, Durham, DH1 3LE, UK\label{addr2a}
           \and
           Georg-August-Universit\"at G\"ottingen, Institut f\"ur Theoretische Physik, Friedrich-Hund-Platz 1, 37077 G\"ottingen, Germany\label{addr3}
}

\date{Received: date / Accepted: date}

\maketitle

\begin{abstract}
  We present the evaluation of the complete set of NLO corrections to three-jet production
  at the LHC. To this end we consider all contributions of ${\cal{O}}(\alphaS^n\alpha^m)$
  with $n+m=3$ and $n+m=4$. This includes in particular also subleading Born contributions of
  electroweak origin, as well as electroweak virtual and QED real-radiative corrections.
  As an application we present results for the three- over two-jet ratio $\Rtt$.  
  While the impact of non-QCD corrections on the total cross section is rather small,
  they can exceed $-10\%$ for high jet transverse momenta. The $\Rtt$ observable turns out
  to be very stable against electroweak corrections, receiving absolute corrections below
  $5\%$ even in the high-$p_T$ region.
\end{abstract}

\section{Introduction}
\label{sec:intro}

Jet-production processes make up the most abundant final states
in hadron-hadron collisions, as carried out at the Large Hadron Collider
(LHC). They are of great importance for the determination of the
strong-coupling constant and provide a central ingredient to
precise determinations of parton density functions (PDFs). At the
same time pure-jet final states constitute promising search grounds for
physics beyond the Standard Model, when looking for resonance peaks
or an excess of events in the tails of transverse-momentum-type distributions. 

Besides being of high phenomenological relevance, jet-production processes
serve as benchmark for various types of perturbative calculations including
fixed-order evaluations, all-orders resummations and parton-shower simulations.
Already the two-jet production channel features quarks and gluons in the
initial and final states and correspondingly various types of spin- and
color-correlations. Beyond the leading order there arise infrared singularities
both in the virtual and real corrections that need to be properly treated.
Further, sensitivity to the actual jet criterion used to define
the cross section emerges. Beyond perturbation theory, there are important corrections
from the fragmentation of final-state partons into hadrons and beam-remnant
interactions such as multi-parton scatterings. 

For hadro-production the next-to-leading order (NLO) QCD corrections are
known to up to five-jet final states~\cite{Ellis:1992en,Giele:1993dj,Nagy:2003tz,Bern:2011ep,Badger:2013yda}.
The computation of the QCD next-to-next-to leading order (NNLO) corrections
to dijet production has recently been completed \cite{Currie:2017eqf}, resulting
in significantly reduced scale uncertainties in the predictions, paving the way
to precision analyses of LHC dijet data. Dedicated studies on the combination of
NLO QCD calculations with parton-shower simulations for dijet production have
been presented in \cite{Alioli:2010xa,Hoche:2012wh}. 

To further improve the theoretical accuracy besides QCD also electroweak (EW) corrections
need to be considered. A first evaluation of the leading weak corrections to
dijet production has been presented in \cite{Dittmaier:2012kx}. These included the
tree-level contributions of ${\cal{O}}(\alpha_s\alpha)$ and ${\cal{O}}(\alpha^2)$ and
weak loop corrections of ${\cal{O}}(\alpha_s^2\alpha)$. Only recently the complete
set of NLO corrections, further including QED virtual and real contributions, was
completed \cite{Frederix:2016ost}. While these corrections are rather small for total
cross sections, they can reach $10-20\%$ for jet transverse momenta in the TeV range. 

A first evaluation of the full set of NLO corrections, of QCD and EW origin, for the
three-jet inclusive cross section has been quoted in \cite{Frederix:2018nkq}. In this
paper we present results for the fully differential calculation of three-jet production
at the LHC to NLO, including all contributions proportional to $\alpha_s^n\alpha^m$
with $n+m=3$ and $n+m=4$. As a first application we consider the observable $\Rtt$,
the ratio of the three-jet and two-jet cross sections, differential in \HTtwo, i.e.
the scalar sum of the two leading-jets transverse momenta. 

Our paper in organised as follows, in Sec.~\ref{sec:setup} we present our calculational
methods and specify our input parameters. In Sec.~\ref{sec:results} we present our
results for the full NLO calculation of the three-jet process and the $\Rtt$ observable
in particular. We give a summary of our findings in Sec.~\ref{sec:conclusions}.

\section{Setup}
\label{sec:setup}

To obtain the results presented in Sec.\ \ref{sec:results} we use 
the \Sherpa Monte-Carlo event generator \cite{Gleisberg:2008ta} 
and interface \cite{Biedermann:2017yoi} it to \Recola\footnote{
  The public version 1.2 of \Recola is used.
} \cite{Actis:2012qn,Actis:2016mpe}.
Therein, the tree-level matrix elements, 
infrared subtractions, process management and phase-space 
integration are provided by \Sherpa for all contributions 
to all processes using its tree-level matrix-element 
generator \Amegic \cite{Krauss:2001iv}. 
It also implements the infrared subtraction \cite{Gleisberg:2007md,Schonherr:2017qcj, 
  Kallweit:2014xda,Kallweit:2015dum,Biedermann:2017yoi,Kallweit:2017khh,
  Chiesa:2017gqx,Greiner:2017mft,Gutschow:2018tuk,Schonherr:2018jva}
in the QCD+QED generalisation of the Catani-Seymour scheme 
\cite{Catani:1996vz,Dittmaier:1999mb,Catani:2002hc,Dittmaier:2008md}, 
including the appropriate initial state mass factorisation counter 
terms. 
\Recola, on the other hand, using the \Collier library \cite{Denner:2016kdg} 
for the evaluation of its scalar and tensor integrals, provides the 
renormalised virtual corrections.

\begin{figure*}
  \centering
  \begin{tabular}{cccc}
  \includegraphics[width=0.18\textwidth]{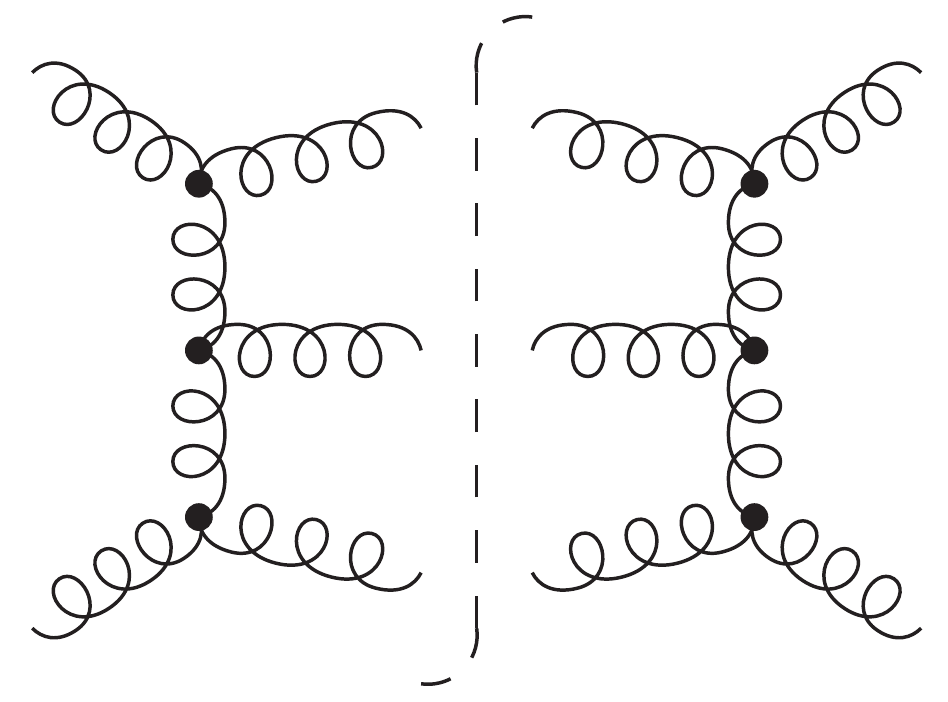} &
  \includegraphics[width=0.18\textwidth]{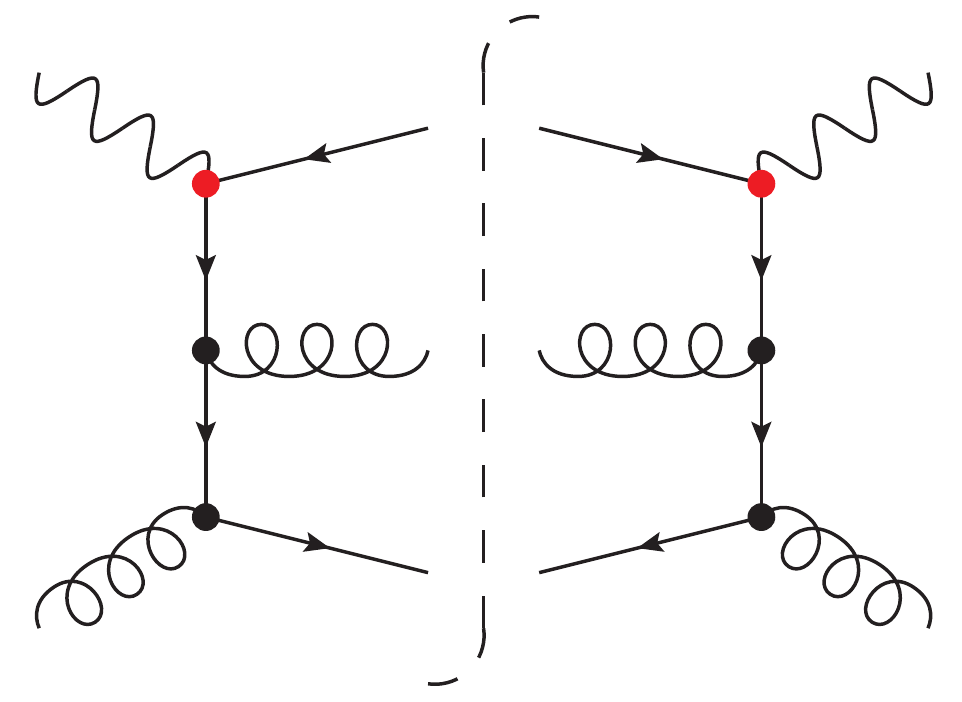} &
  \includegraphics[width=0.18\textwidth]{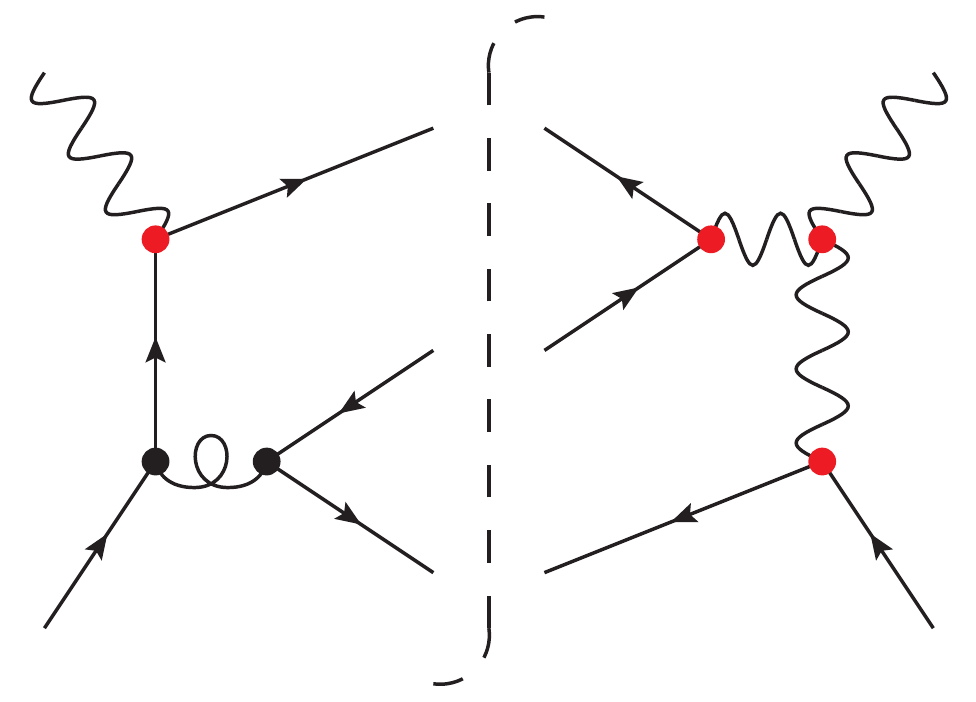} &
  \includegraphics[width=0.18\textwidth]{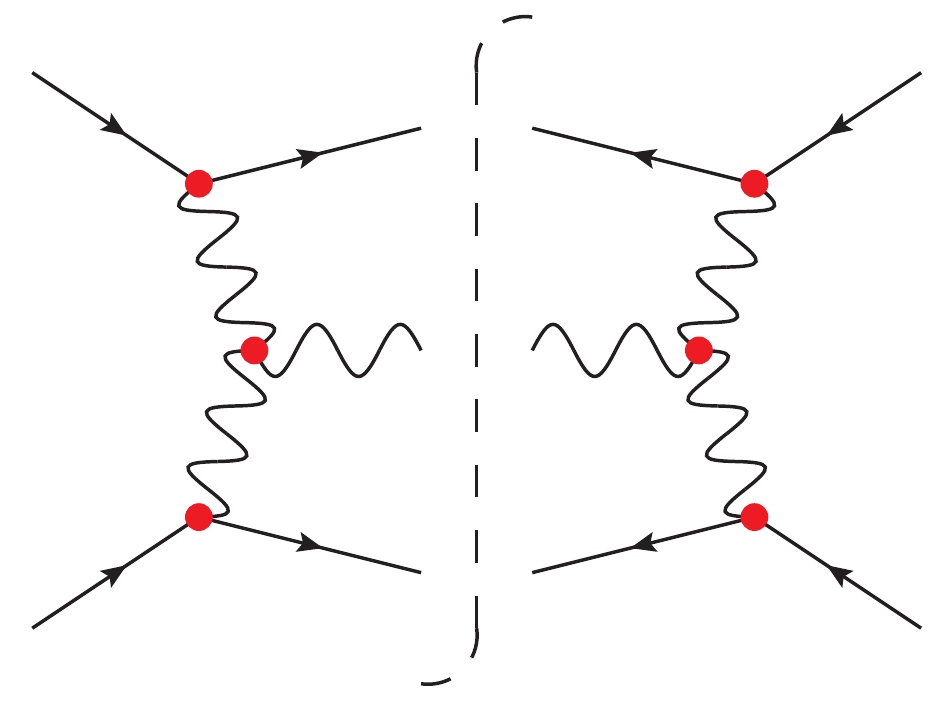} \\
  \includegraphics[width=0.18\textwidth]{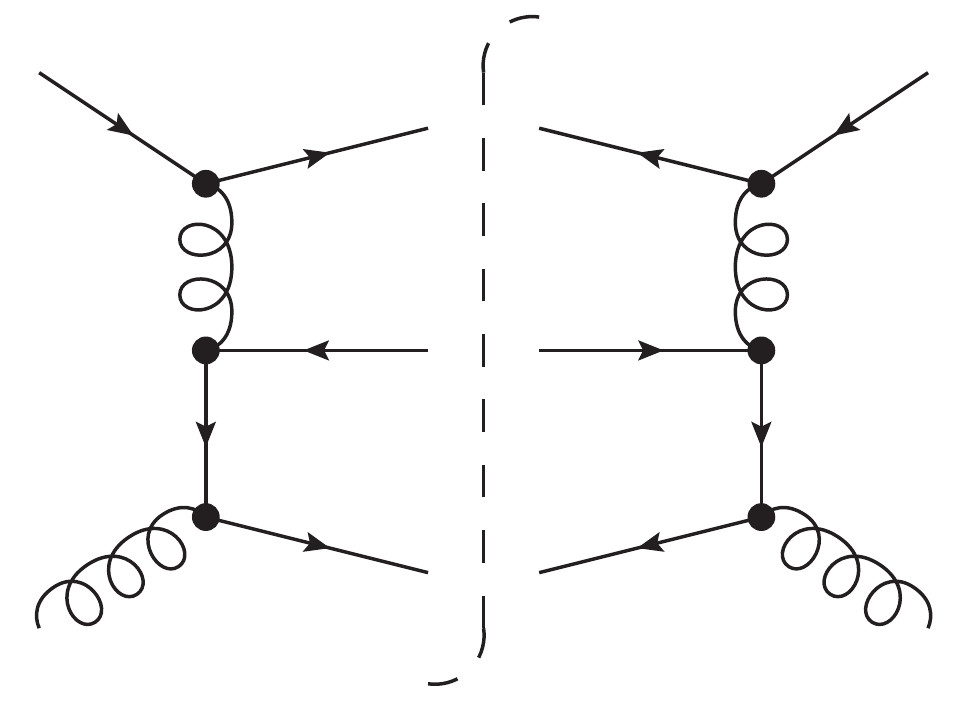} &
  \includegraphics[width=0.18\textwidth]{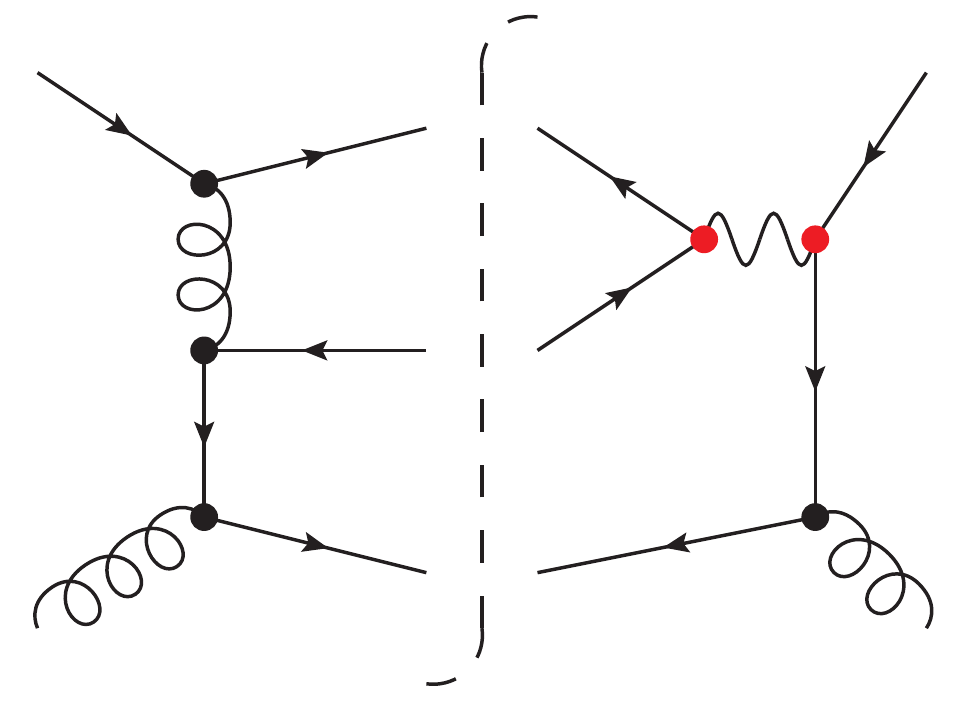} &
  \includegraphics[width=0.18\textwidth]{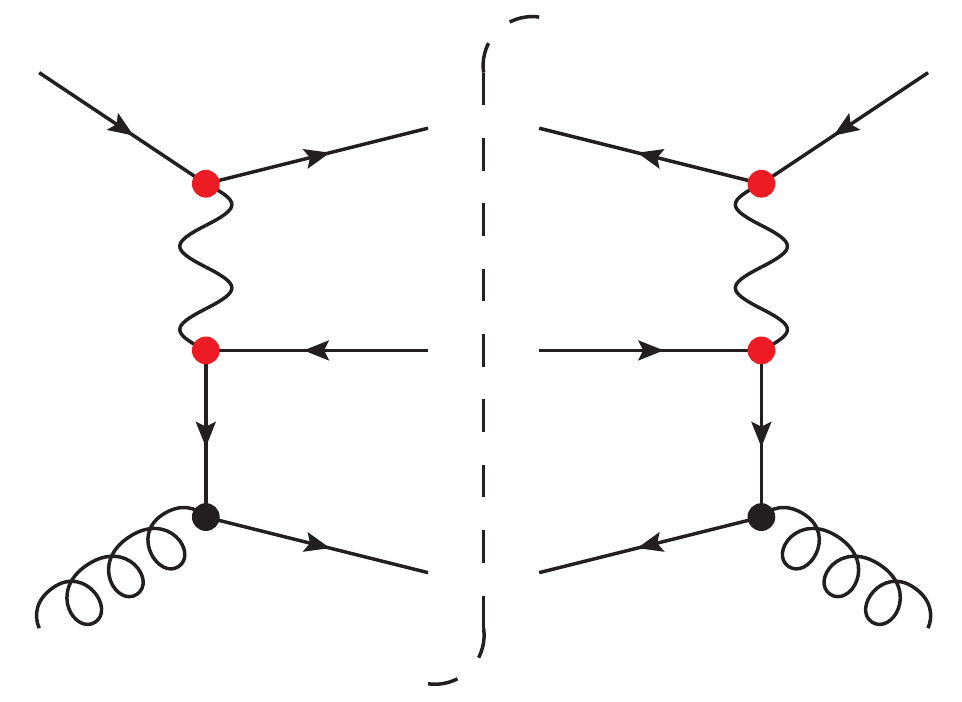} &
  \includegraphics[width=0.18\textwidth]{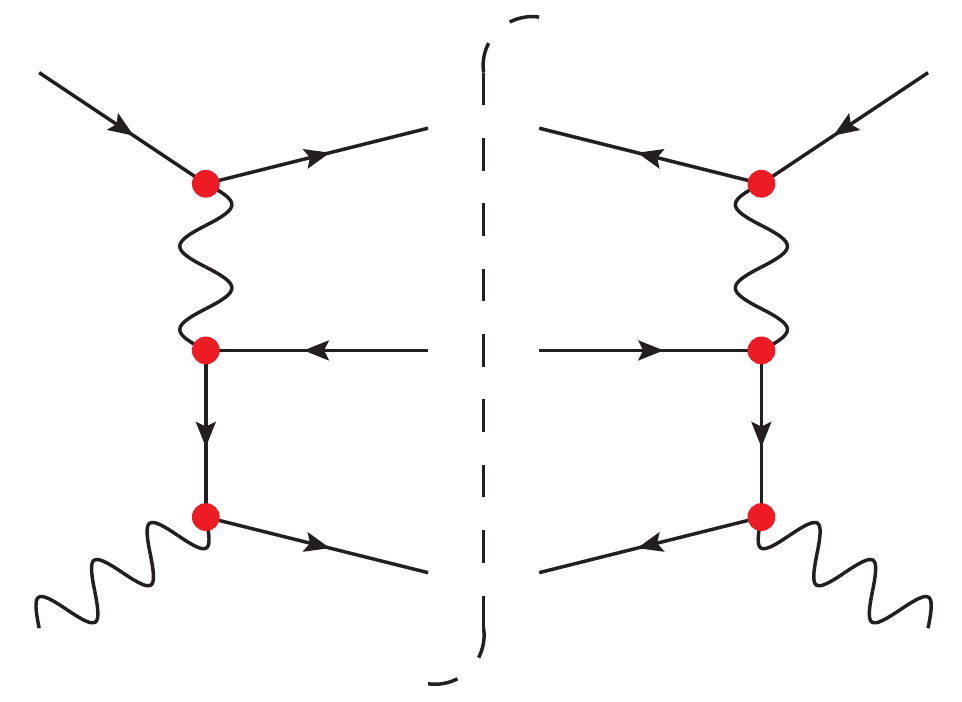} \\
  $\order(\alphaS^3)$ &
  $\order(\alphaS^2\alpha)$ &
  $\order(\alphaS\alpha^2)$ &
  $\order(\alpha^3)$
  \end{tabular}\\
  \caption{
    Representative leading and subleading tree-level diagrams for 
    $\mathrm{pp}\to 3j$ production. The occurrence of QCD and
    electroweak interferences, internal electroweak bosons and external
    photons (wavy lines) in the initial and final state are exemplified.
    While QCD vertices are marked by a black dot, EW interactions are
    indicated in red.
    \label{fig:diagrams-LO}
  }
\end{figure*}

\begin{figure*}
  \centering
  \begin{tabular}{ccccc}
  \includegraphics[width=0.18\textwidth]{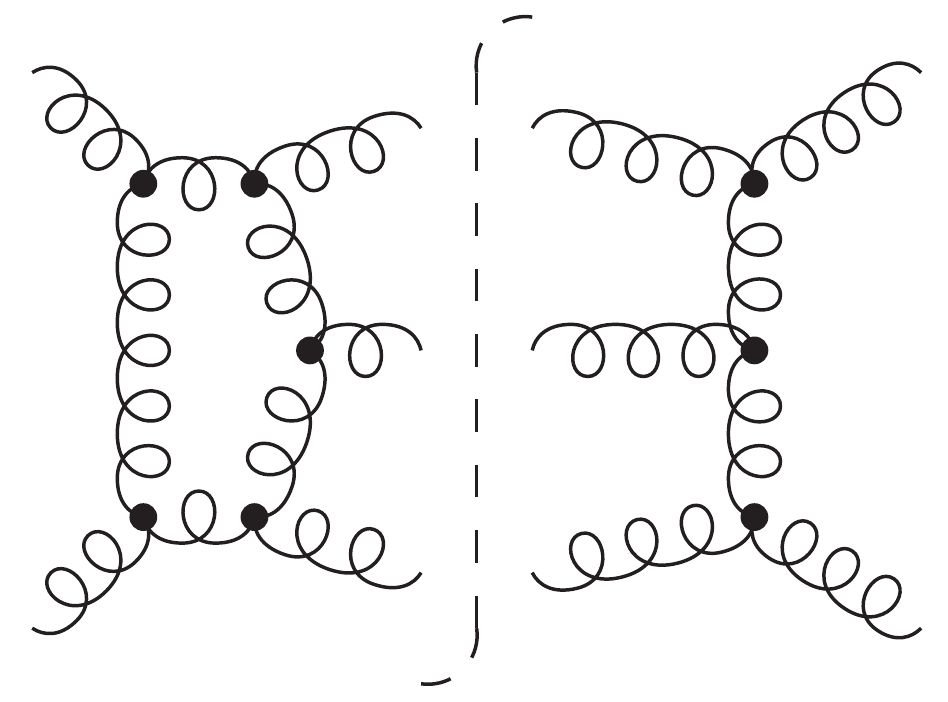} &
  \includegraphics[width=0.18\textwidth]{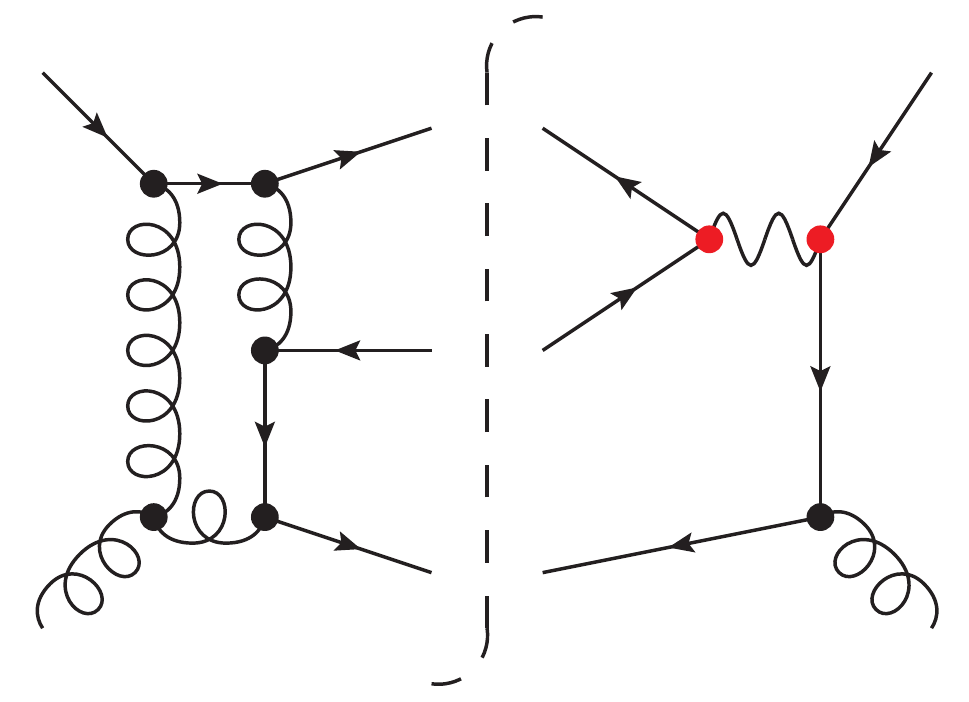} &
  \includegraphics[width=0.18\textwidth]{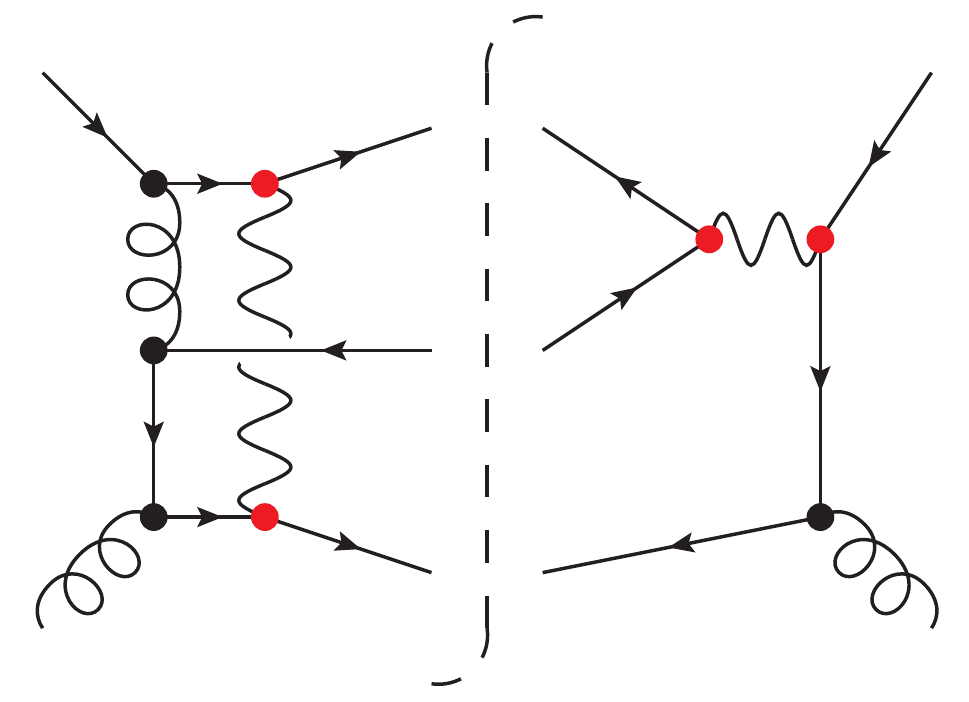} &
  \includegraphics[width=0.18\textwidth]{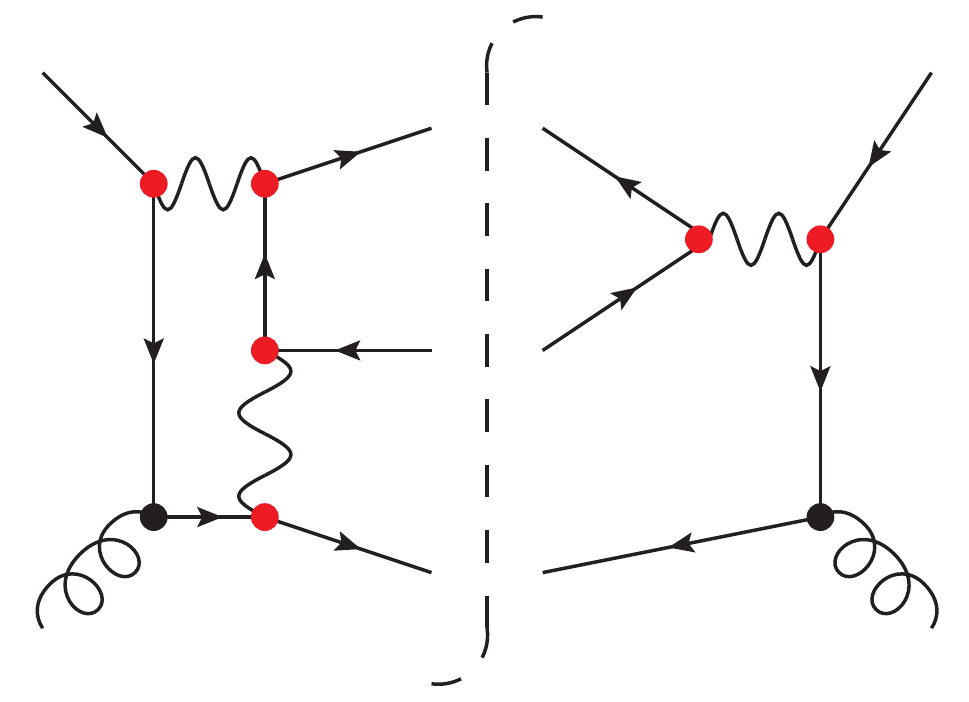} &
  \includegraphics[width=0.18\textwidth]{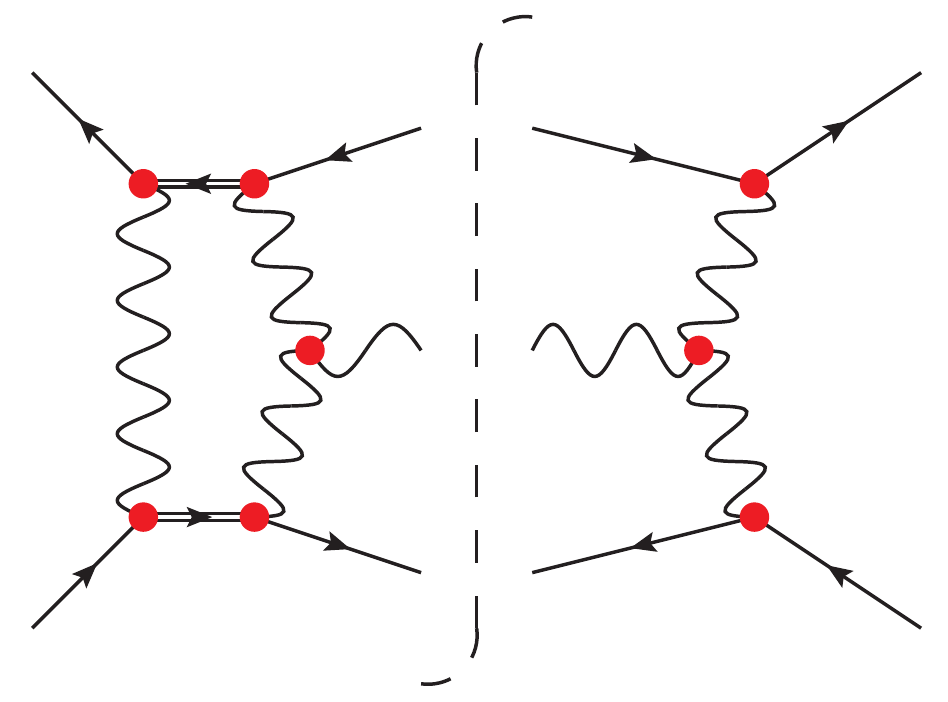} \\
  \includegraphics[width=0.18\textwidth]{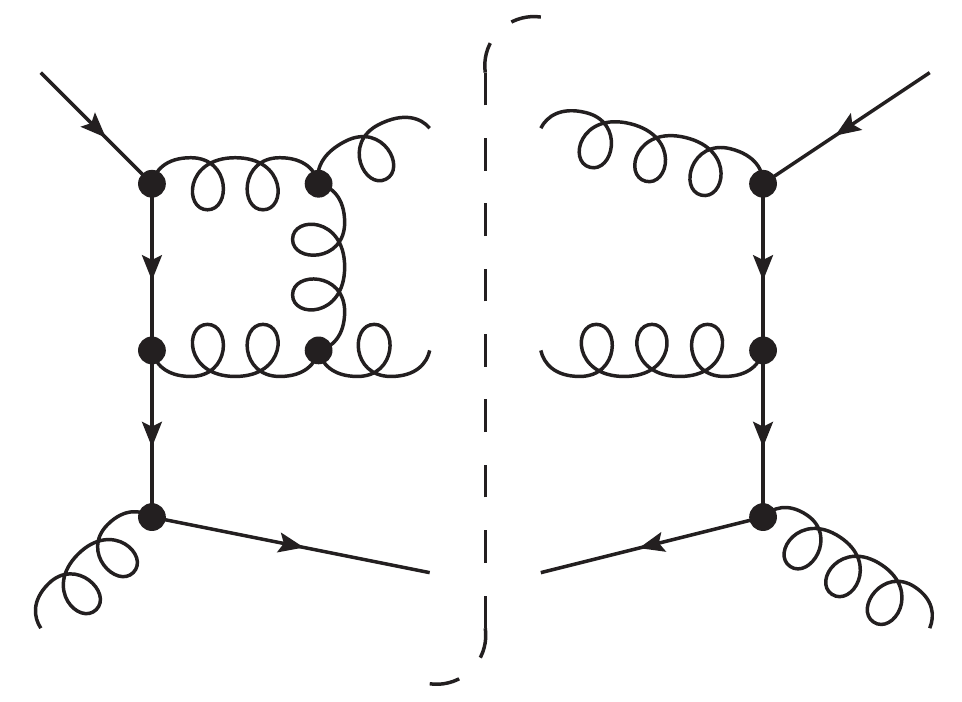} &
  \includegraphics[width=0.18\textwidth]{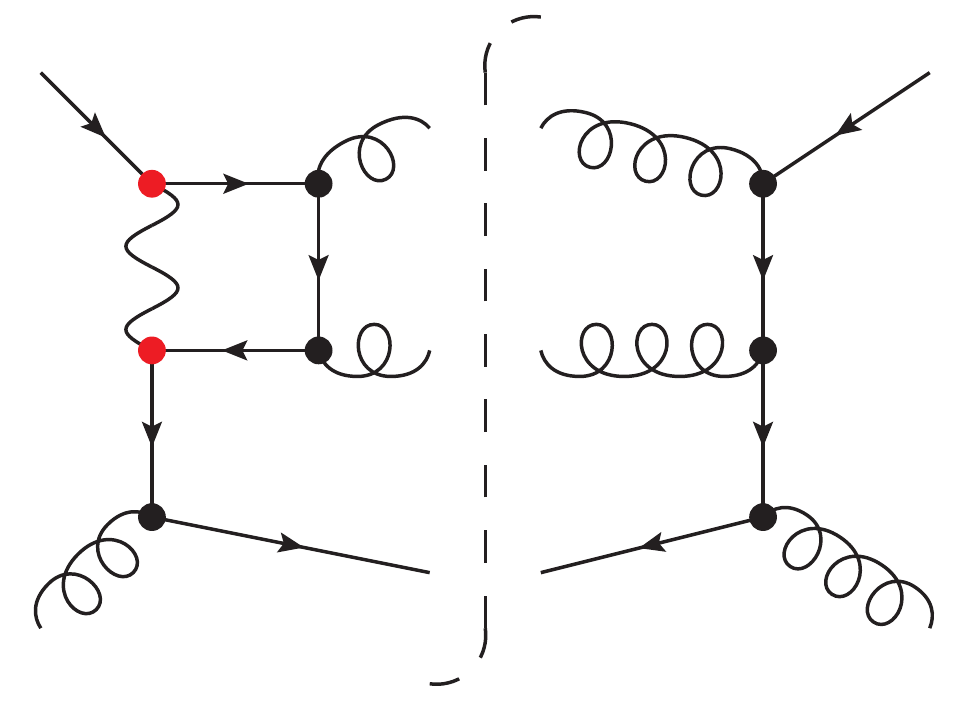} &
  \includegraphics[width=0.18\textwidth]{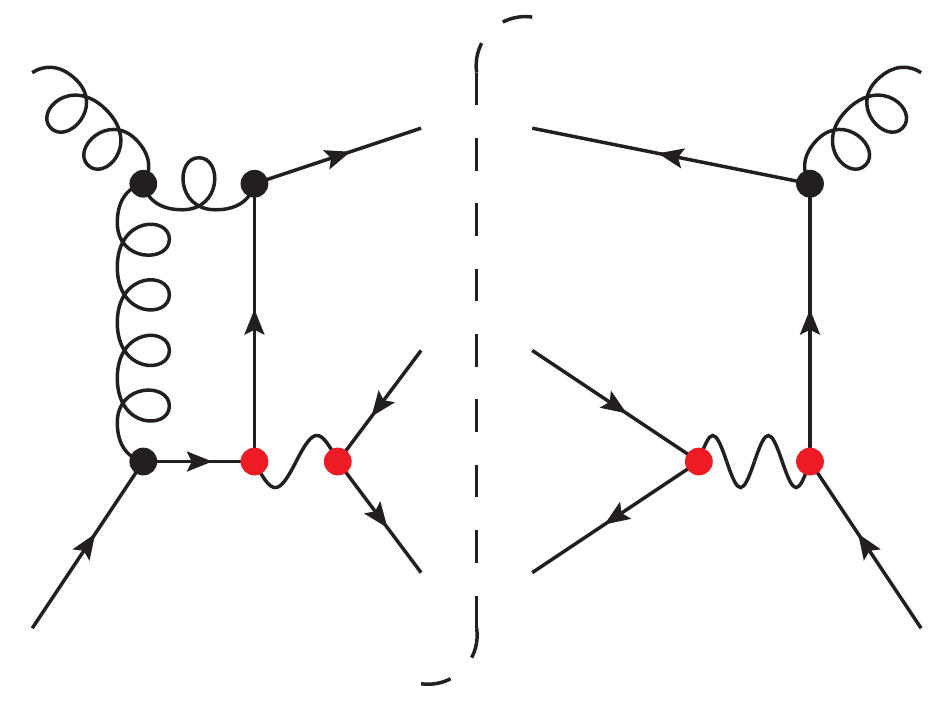} &
  \includegraphics[width=0.18\textwidth]{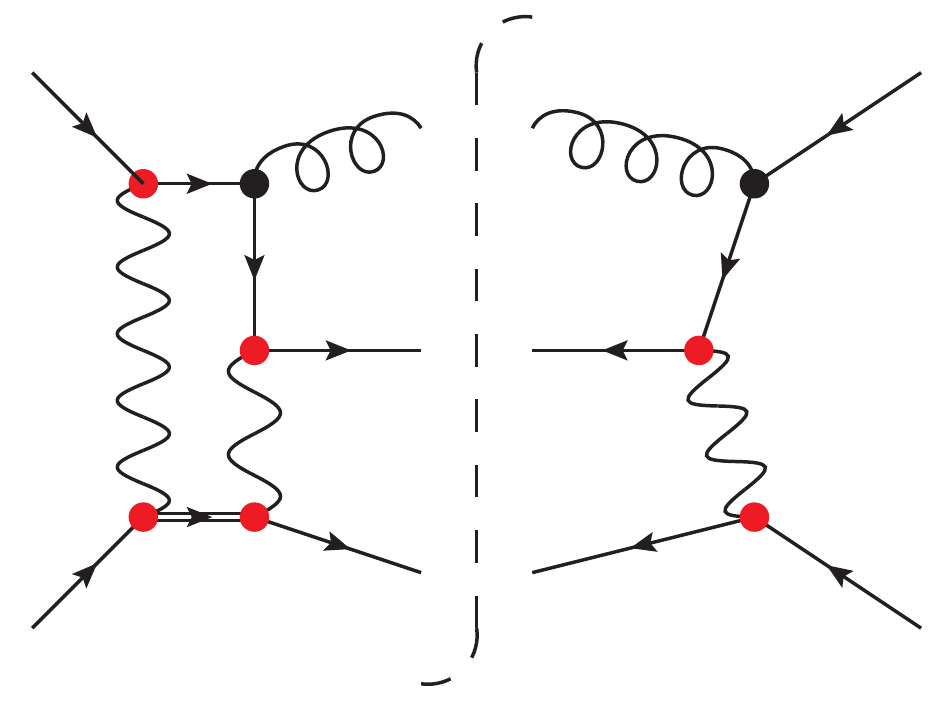} &
  \includegraphics[width=0.18\textwidth]{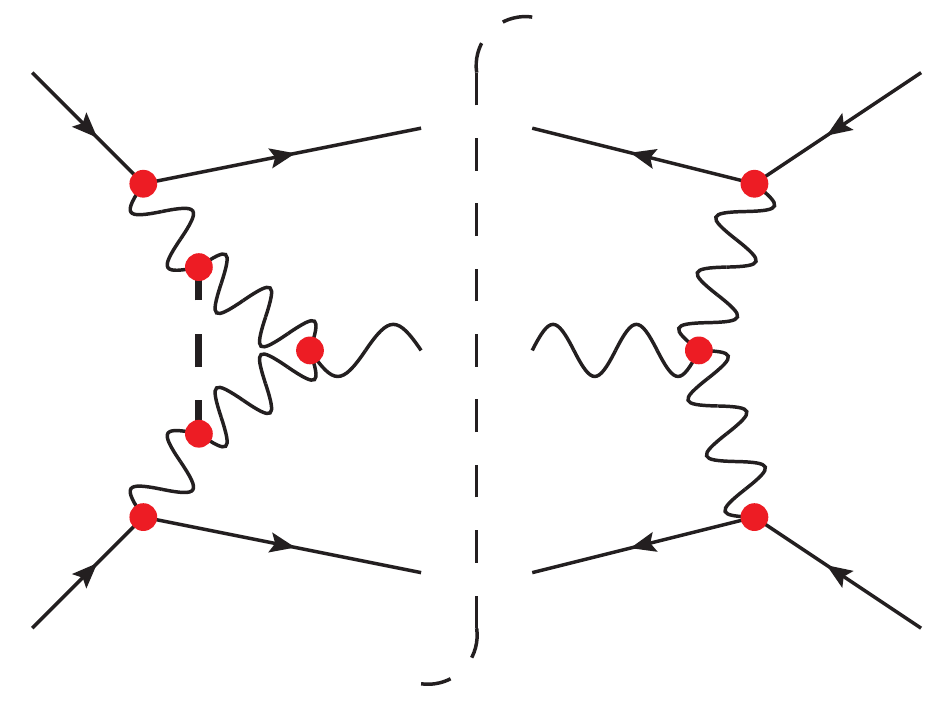} \\
  $\order(\alphaS^4)$ &
  $\order(\alphaS^3\alpha)$ &
  $\order(\alphaS^2\alpha^2)$ &
  $\order(\alphaS\alpha^3)$ &
  $\order(\alpha^4)$
  \end{tabular}
  \caption{
    Representative leading and subleading virtual correction diagrams for 
    $\mathrm{pp}\to 3j$ production. The occurrence of interferences, 
    QCD and EW loops, gauge boson (wavy line), Higgs boson (dashed line) and
    top quark (double line) exchange as well as external photons are exemplified.
    While QCD vertices are marked by a black dot, EW interactions are indicated in red.
    \label{fig:diagrams-VIRT}
  }
\end{figure*}

All calculations are performed in the framework of the Standard Model,
assuming a diagonal CKM matrix, and using the five-flavour scheme, 
i.e.\ treating the bottom quark as massless. 
The complex mass scheme \cite{Denner:2005fg,Denner:2014zga} is 
used to consistently treat intermediate resonances in the 
contributing amplitudes.
All electroweak Standard Model parameters are defined in the $\Gmu$-scheme, 
and virtual amplitudes are renormalised correspondingly. 
Consequently, the following set of input parameters is used throughout
\begin{center}
  \begin{tabular}{rclrcl}
    $\Gmu$ &\shortequal& $1.16639\times 10^{-5}\; \text{GeV}^{-2}$ &&& \\
    $m_W$ &\shortequal& $80.385\; \text{GeV}$  & $\Gamma_W$ &\shortequal& $2.085\; \text{GeV}$ \\
    $m_Z$ &\shortequal& $91.1876\; \text{GeV}$ & $\Gamma_Z$ &\shortequal& $2.4952\; \text{GeV}$ \\
    $m_h$ &\shortequal& $125.0\; \text{GeV}$   & $\Gamma_h$ &\shortequal& $0.00407\; \text{GeV}$\\
    $m_t$ &\shortequal& $173.21\; \text{GeV}$   & $\Gamma_t$ &\shortequal& $1.3394$\; \text{GeV}\,.
  \end{tabular}
\end{center}
All other masses and widths are set to zero. 
In the above, 
\begin{equation}
  \label{eq:defalpha}
  \begin{split}
    \alpha
    \,=&\;
      \left|
	\frac{\sqrt{2}\;\Gmu\;\mu^2_W\,\sin^2\theta_\text{w}}{\pi}
    \right|\,,
  \end{split}
\end{equation}
defines the electromagnetic coupling. 
The complex mass of particle $i$ and the weak mixing angle 
are defined according to
\begin{equation}
  \mu_i^2=m_i^2-\mathrm{i}m_i\Gamma_i
  \qquad\text{and}\qquad
  \sin^2\theta_\text{w}=1-\frac{\mu_W^2}{\mu_Z^2}\;,
\end{equation}
respectively.

For the parton density functions we use the NNPDF3.1 NLO PDF set
\cite{Bertone:2017bme} with $\alphaS(m_Z)=0.118$ and 
including QED effects (at $\order(\alpha)$, $\order(\alphaS\alpha)$ and 
$\order(\alpha^2)$) in the parton evolution employing the LUXqed scheme 
\cite{Manohar:2016nzj,Manohar:2017eqh} \footnote{
  To be precise the \texttt{NNPDF31\_nlo\_as\_0118\_luxqed} PDF set is used.
}.
They are interfaced through \LHAPDF \cite{Buckley:2014ana}.
The renormalisation and factorisation scales are defined as
\begin{equation}\label{eq:muRF}
  \begin{split}
    \mu_R\,=\,\mu_F\,=\,\HThalf\;.
  \end{split}
\end{equation}
The variable $\HT$ is thereby given by the scalar sum of all final-state
particles' transverse momenta without applying any jet clustering. 
To estimate the uncertainty on our computation
from uncalculated higher-order contributions, we vary the renormalisation and 
factorisation scales independently by the customary factor two, 
keeping $\tfrac{1}{2}\leq\mu_R/\mu_F\leq 2$. 
All scale variations were calculated on-the-fly using the event-reweighting
algorithm detailed in \cite{Bothmann:2016nao}. 

\section{Results}
\label{sec:results}

In this section numerical results for the production of a three-jet 
final state at next-to-leading order accuracy in proton-proton collisions at
a centre-of-mass energy of 13\,TeV are presented. 
We generate the respective matrix elements at all contributing orders 
for all partonic processes with massless three (Born and virtual corrections) 
and four body final states (real corrections). As final-state particles
we consider five quark flavours and gluons, as well as photons,
leptons and neutrinos. Jets are then defined through the anti-$k_t$ algorithm 
\cite{Cacciari:2008gp} using \Fastjet \cite{Cacciari:2011ma}, 
with $R=0.4$ as radial parameter. 
All massless particles of our calculation, except for the neutrinos, 
are considered as jet constituents. 
Jets with a net lepton number\footnote{
  A jet with a lepton and an anti-lepton, if they are of the same lepton 
  flavour, has net lepton number zero.
} and within $|\eta|<2.5$ are removed from the list of jets. 
The final state then has to contain at least three surviving jets with 
$|\eta(j)|<2.8$, of which the leading jet, ordered in transverse 
momentum, must have $\pT(j_1)>80\,\GeV$ and all subleading jets 
$\pT(j_i)>60\,\GeV$ ($i>1$). 
This ensures that a jet definition with inherent lepton rejection, 
which is both infrared-safe at NLO and close to 
experimental analysis strategies, is used. 
Nonetheless, it is worth pointing out that lepton final states may survive 
this lepton-anti-tagged jet definition if either a collinear lepton pair 
is contained in a single jet (possibly coming from a collinear 
$\gamma\to\ell^+\ell^-$ splitting), or the jet containing the lepton is 
outside the rapidity range in which the lepton can be identified. 
To analyse our results we use the \Rivet\ package \cite{Buckley:2010ar}.

The full NLO $n$-jet production cross section can be decomposed into
contributions of varying power of the strong and electromagnetic coupling.
In what follows we employ the convention:
\begin{equation}
  \label{eq:xsec_contribs}
  \begin{split}
    \begin{array}{rclrcl}
      \sigma_{nj}
      & = & \sigma_{nj}^\LO + \sigma_{nj}^{\DNLO}, &&&
      \\[2mm]
      \sigma_{nj}^\LO
      & = & \sum\limits_{i=0}^{n} \sigma^{\LO_{i}}_{nj},
      &
      \order \left( \sigma^{\LO_{i}}_{nj} \right)
      & = & \alphaS^{n-i} \alpha^{i},
      \\[2mm]
      \sigma_{nj}^{\DNLO}
      & = & \sum\limits_{i=0}^{n+1} \sigma^{\DNLO_{i}}_{nj},
      &
      \order \left( \sigma^{\DNLO_{i}}_{nj} \right)
      & = & \alphaS^{n+1-i} \alpha^{i},
    \end{array}
  \end{split}
\end{equation}
such that $\DNLO_i$ accounts for the virtual and real
\emph{QCD} corrections while $\DNLO_{i+1}$ accounts for the virtual and real
\emph{electroweak} corrections to $\LO_i$. Representative diagrams
for the various tree-level and virtual contributions can be found
in Fig.~\ref{fig:diagrams-LO} and Fig.~\ref{fig:diagrams-VIRT}, respectively.
It is worth noting that our full NLO calculation in the five-flavour scheme
is indeed sensitive to the full Standard Model spectrum,
including the top-quark, the Higgs boson and all lepton and neutrino flavours.

Based on the above decomposition we can furthermore define the pure QCD LO and NLO
cross sections as
\begin{equation}
  \label{eq:phys_xsec_qcd}
  \begin{split}
    \sigma_{nj}^{\LO\;\QCD}&=\sigma_{nj}^{\LOzero},\\
    \sigma_{nj}^{\NLO\;\QCD}&=\sigma_{nj}^{\LOzero}+\sigma_{nj}^{\DNLOzero},
    \end{split}
\end{equation}
respectively. The pure NLO EW corrections and their additive and
multiplicative combination with the QCD process are defined as
\begin{equation}
  \label{eq:phys_xsec_ew}
  \begin{split}
    \sigma_{nj}^{\NLO\;\EW}
    &=\sigma_{nj}^{\LOzero}+\sigma_{nj}^{\DNLOone},
    \vphantom{\frac{\sigma_{nj}^{\DNLOzero}}{\sigma_{nj}^{\LOzero}}}
    \\
    \sigma_{nj}^{\NLO\;\QCDpEW}
    &=\sigma_{nj}^{\LOzero}+\sigma_{nj}^{\DNLOzero}+\sigma_{nj}^{\DNLOone},
    \\
    \sigma_{nj}^{\NLO\;\QCDtEW}
    &=\sigma_{nj}^{\LOzero}
      \left(1+\frac{\sigma_{nj}^{\DNLOzero}}{\sigma_{nj}^{\LOzero}}\right)
      \left(1+\frac{\sigma_{nj}^{\DNLOone}}{\sigma_{nj}^{\LOzero}}\right).
    \end{split}
    \hspace*{-10mm}
\end{equation}
The difference between the additive and multiplicative combination provides
an estimate of uncalculated mixed QCD-EW NNLO corrections of $\order(\alphaS\alpha)$, wrt.\ LO QCD.

\begin{table*}[t!]
  \centering
  \begin{tabular}{l||c||c|c|c|c||c|c|c|c|c}
    \Hl & \NLO      & $\frac{\LOzero}{\NLO}$ & $\frac{\LOone}{\NLO}$ 
                    & $\frac{\LOtwo}{\NLO}$ & $\frac{\LOthree}{\NLO}$ 
                    & $\frac{\DNLOzero}{\NLO}$ & $\frac{\DNLOone}{\NLO}$ 
                    & $\frac{\DNLOtwo}{\NLO}$ & $\frac{\DNLOthree}{\NLO}$ 
                    & $\frac{\DNLOfour}{\NLO}$ \\
                    & [nb]
                    & [\%]
                    & [\%]
                    & [\%]
                    & [\%]
                    & [\%]
                    & [\%]
                    & [\%]
                    & [\%]
                    & [\%]
                    \\\hline\hline
    $\sigma_{2j}$\hl & $3385(3)$ & $67.34(6)$
                     & $0.0713(1)$
                     & $0.03915(4)$
                     & -- & $32.59(8)$
                     & $-0.118(7)$
                     & $0.0759(3)$
                     & $0.00022(1)$
                     & -- \\\hline
    $\sigma_{3j}$\hl & $169(1)$ & $148(1)$
                     & $0.293(2)$
                     & $0.196(2)$
                     & $0.00217(2)$
                     & $-48.4(8)$
                     & $-0.74(1)$
                     & $0.344(7)$
                     & $-0.00433(6)$
                     & $0.0135(2)$
    \end{tabular}
  \caption{
    Full NLO fiducial cross section for two- and three-jet production in the phase space detailed in the
    text, i.e. $\pT(j_1)>80\; \GeV$ and $\pT(j_i)>60\; \GeV$ ($i>1$). Besides the total cross section the
    relative contributions for the terms specified in Eqs.~(\ref{eq:xsec_contribs}) are given.
    \label{tab:xs}
  }
\end{table*}

\begin{table*}[t!]
  \centering
  \begin{tabular}{l||c||c|c|c|c||c|c|c|c|c}
    \Hl & \NLO      & $\frac{\LOzero}{\NLO}$ & $\frac{\LOone}{\NLO}$ 
                    & $\frac{\LOtwo}{\NLO}$ & $\frac{\LOthree}{\NLO}$ 
                    & $\frac{\DNLOzero}{\NLO}$ & $\frac{\DNLOone}{\NLO}$ 
                    & $\frac{\DNLOtwo}{\NLO}$ & $\frac{\DNLOthree}{\NLO}$ 
                    & $\frac{\DNLOfour}{\NLO}$ \\
                    & [fb]
                    & [\%]
                    & [\%]
                    & [\%]
                    & [\%]
                    & [\%]
                    & [\%]
                    & [\%]
                    & [\%]
                    & [\%]
                    \\\hline\hline
    $\sigma_{2j}$\hl & $51.9(6)$ & $60(1)$
                     & $7.07(8)$
                     & $1.82(2)$
                     & -- & $36.9(8)$
                     & $-4.5(1)$
                     & $-1.02(2)$
                     & $-0.552(7)$
                     & -- \\\hline
    $\sigma_{3j}$\hl & $40.0(4)$ & $99(1)$
                     & $8.6(1)$
                     & $2.05(4)$
                     & $0.061(1)$
                     & $-0.9(9)$
                     & $-9.8(4)$
                     & $1.09(7)$
                     & $0.057(4)$
                     & $0.314(5)$
    \end{tabular}
  \caption{
    As Table~\ref{tab:xs} but with the additional requirement of $\pT(j_1)>2\; \TeV$.  
    \label{tab:xs_2tev}
  }
\end{table*}

\begin{table*}[t!]
  \centering
  \begin{tabular}{l||c||c|c|c|c}
    \Hl & \NLO      & \LO\;\QCD & \NLO\;\QCD
                    & \NLO\;\EW
                    & \NLO\;\QCDpEW \\
                    & [nb]
                    & [nb]
                    & [nb]
                    & [nb]
                    & [nb]
                    \\\hline\hline
    $\sigma_{2j}$\hl & $3385(3)^{+334}_{-338}$
                     & $2279.4(6)^{+553.7}_{-404.4}$
                     & $3383(3)^{+335}_{-338}$
                     & $2275.4(6)^{+552.4}_{-403.5}$
                     & $3379(3)^{+333}_{-338}$
                     \\\hline
    $\sigma_{3j}$\hl & $169(1)^{+16}_{-73}$
                     & $249.86(6)^{+102.28}_{-67.89}$
                     & $168(1)^{+16}_{-73}$
                     & $248.62(6)^{+101.62}_{-67.46}$
                     & $167(1)^{+17}_{-73}$

    \end{tabular}
  \caption{
    Fiducial cross sections for two- and three-jet production and their corresponding scale uncertainties for a
    leading-jet selection of $\pT(j_1)>80\; \GeV$. The respective cross section definitions are given in
    Eqs.~(\ref{eq:phys_xsec_qcd}) and~(\ref{eq:phys_xsec_ew}).
    \label{tab:xs_phys}
  }
\end{table*}

\begin{table*}[t!]
  \centering
  \begin{tabular}{l||c||c|c|c|c}
    \Hl & \NLO      & \LO\;\QCD & \NLO\;\QCD
                    & \NLO\;\EW
                    & \NLO\;\QCDpEW \\
                    & [fb]
                    & [fb]
                    & [fb]
                    & [fb]
                    & [fb]
                    \\\hline\hline
    $\sigma_{2j}$\hl & $51.9(6)^{+5.9}_{-6.7}$
                    & $31.2(5)^{+11.4}_{-7.9}$
                    & $50.4(6)^{+7.1}_{-7.3}$
                    & $28.9(5)^{+9.6}_{-6.7}$
                    & $48.1(6)^{+5.2}_{-6.1}$
                    \\\hline
    $\sigma_{3j}$\hl & $40.0(4)^{+0.4}_{-6.9}$
                    & $39.4(2)^{+19.0}_{-12.1}$
                    & $39.0(4)^{+0.0}_{-5.0}$
                    & $35.5(2)^{+15.7}_{-10.2}$
                    & $35.1(4)^{+0.9}_{-8.2}$
    \end{tabular}
  \caption{
    As Table~\ref{tab:xs_phys} but with the additional requirement of $\pT(j_1)>2\; \TeV$.  
    \label{tab:xs_phys_2tev}
  }
\end{table*}

We start our discussion of results by listing the inclusive two- and three-jet
cross sections for leading-jet selections of $\pT(j_1)>80\,\GeV$ and $\pT(j_1)>2\,\TeV$
in Tables \ref{tab:xs} and \ref{tab:xs_2tev}, respectively.
We quote results at full NLO accuracy in the Standard Model and list their decomposition into all contributing
orders. The numbers quoted in parantheses indicate the statistical error estimate on the last digit given.
For a leading jet requirement of $\pT(j_1)>80\,\GeV$ corrections of EW origin are
generally rather small, reaching for the three-jet case at most a relative contribution to the
full NLO result of $-0.7\%$ for $\DNLOone$. The dominant corrections are of QCD nature and account for
$+33\%$ and $-48\%$ for two- and three-jet production, respectively. 

Requiring $\pT(j_1)>2\,\TeV$
changes the picture. While for the two-jet process the QCD NLO corrections are
still dominating, amounting to $+37\%$, QCD-EW mixed Born and EW one-loop
contributions clearly become sizeable, though they largely cancel. For
three-jet production in this selection and scale choice the NLO QCD 
corrections are, accidentally, miniscule, below $-1\%$.
However, the Born contributions of EW origin reach a total of $+11\%$ but largely
get cancelled by the $\DNLOone$ terms that contribute $-10\%$ to the total
NLO result. 

In Tables \ref{tab:xs_phys} and \ref{tab:xs_phys_2tev} we quote two- and
three-jet cross sections at full NLO, LO QCD, NLO QCD, NLO EW and \NLO\;\QCDpEW\
for the leading-jet selections of $\pT(j_1)>80\,\GeV$ and $\pT(j_1)>2\,\TeV$, respectively.
Besides the nominal cross sections we give their scale uncertainty estimates obtained from
7-point variations around the central scale choice $\mu_R\,=\,\mu_F\,=\,\HThalf$.
A significant reduction in particular of the upward variations wrt. LO QCD is observed for
predictions including the $\DNLOzero$ terms. Adding the $\DNLOone$ corrections, however, has
no sizeable effect on the scale uncertainties. Furthermore, no systematic reduction of the
scale uncertainties of the full NLO results in comparison to the \NLO\;\QCDpEW\ predictions
is observed.

In principle, the addition of a $\pT>2\,\text{TeV}$ requirement on the leading 
jet, while leaving the subleading jets at $\pT>60\,\text{GeV}$ only, introduces a
large scale hierarchy to cross section results presented in Tables 
\ref{tab:xs_2tev} and \ref{tab:xs_phys_2tev}. In principle, this mandates the inclusion
of a resummation of the corresponding potentially large logarithms. However, no perturbative
instabilities were encountered in this region and we, thus, consider the results reliable. 
Similar considerations, of course, also apply to the tails of the distributions shown in
the following.

\begin{figure*}[t!]
  \centering
  \includegraphics[width=.45\linewidth]{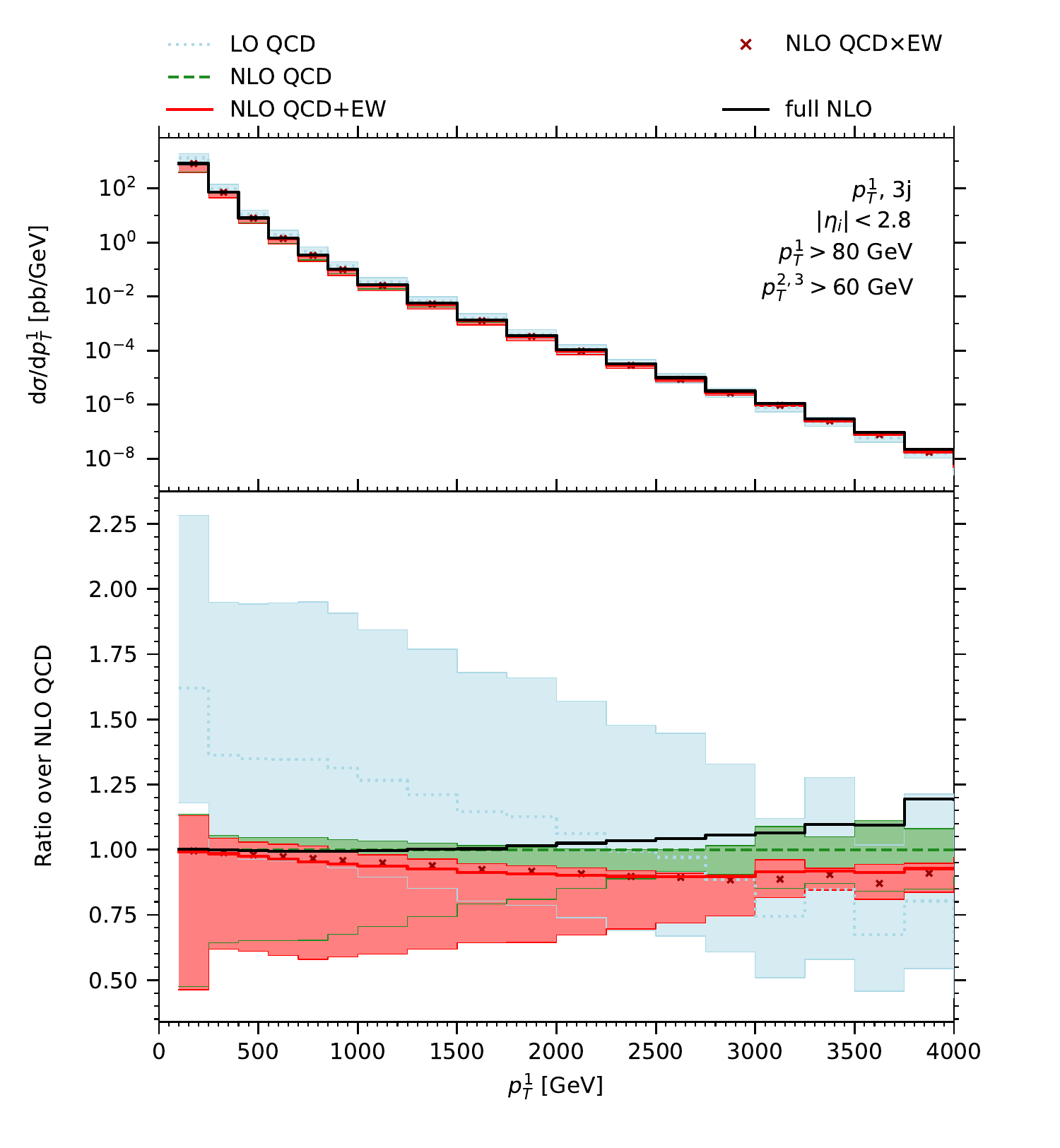}
  \includegraphics[width=.45\linewidth]{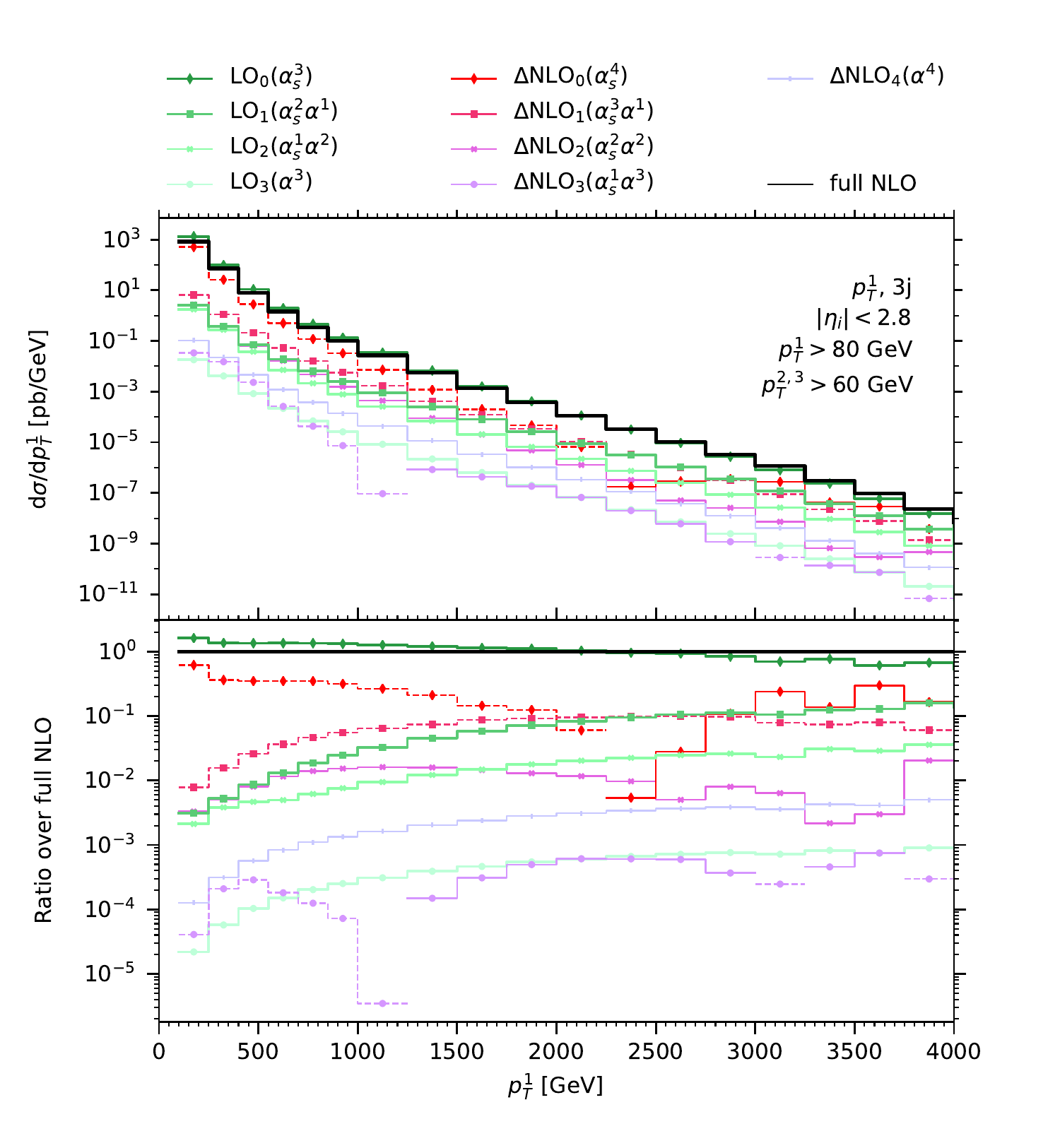}
  \caption{
    Leading jet transverse momentum in three-jet production.
    Left: Theoretical uncertainties at LO, NLO QCD, NLO QCD+EW and full NLO.
    Right: Decomposition of the full NLO result in its contributions defined in Eqs.~(\ref{eq:xsec_contribs}).
    \label{fig:pT1}
  }
\end{figure*}

\begin{figure*}[t!]
  \centering
  \includegraphics[width=.45\linewidth]{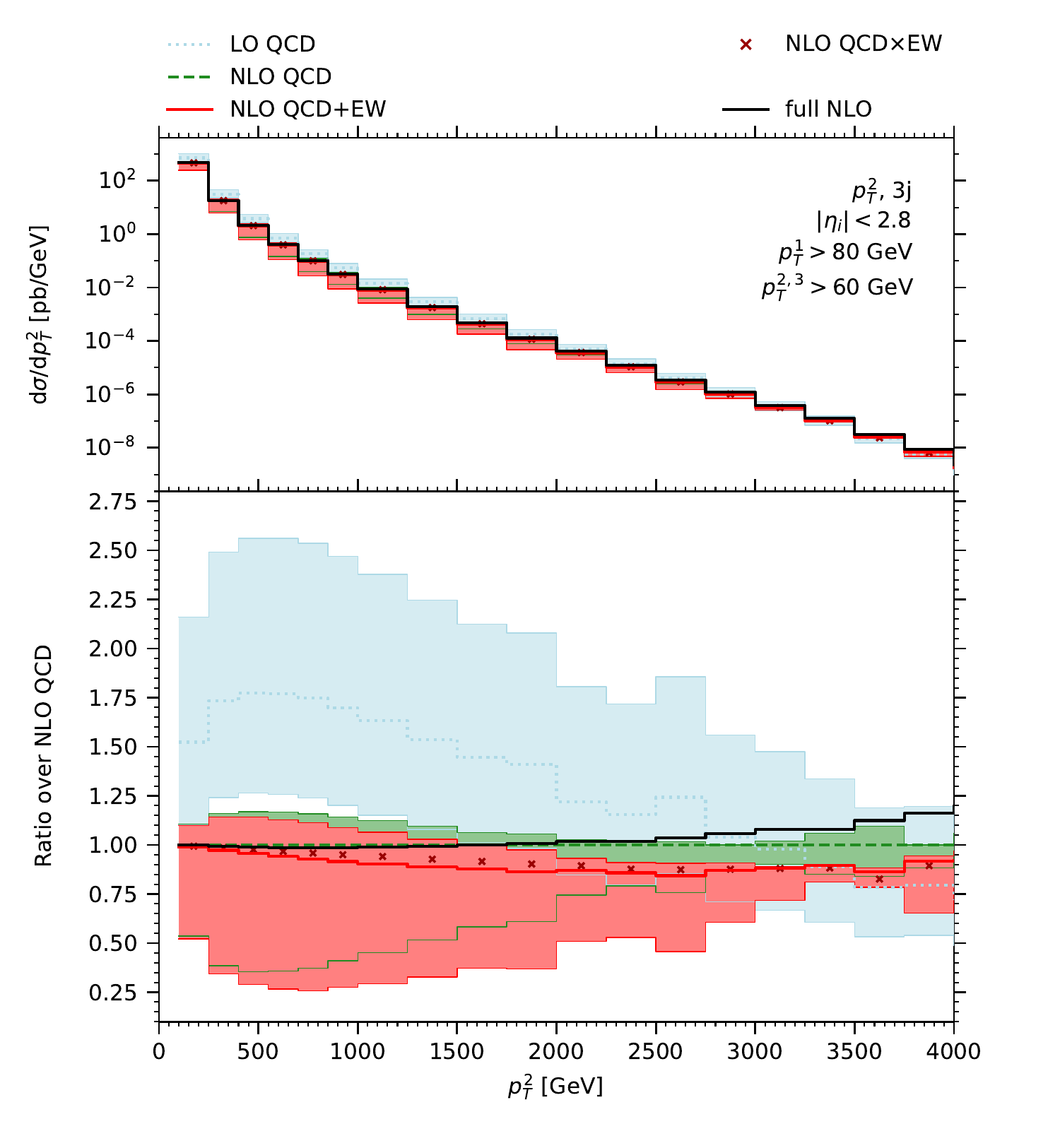}
  \includegraphics[width=.45\linewidth]{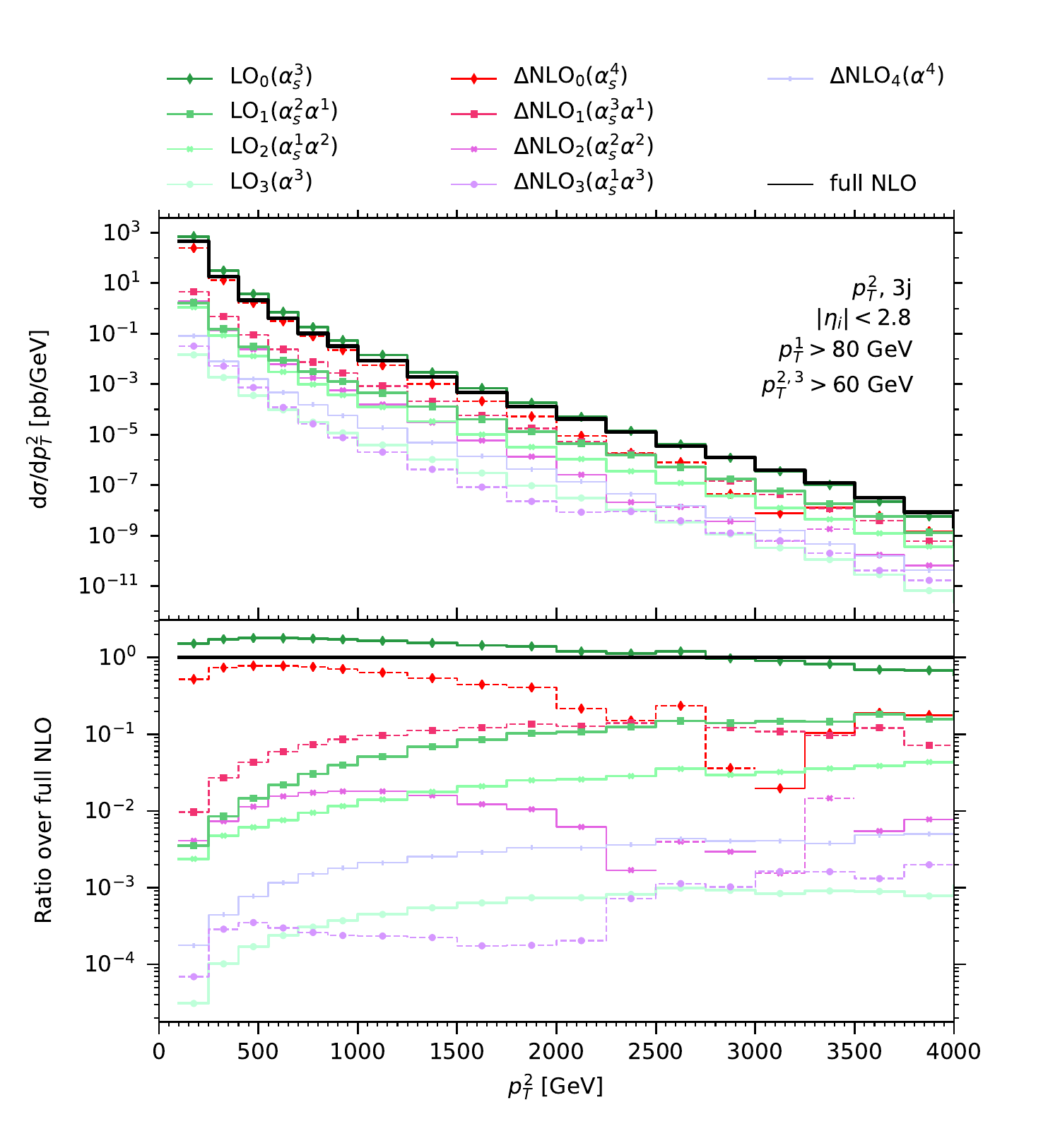}
  \caption{
    Subleading jet transverse momentum in three-jet production.
    Left: Theoretical uncertainties at LO, NLO QCD, NLO QCD+EW and full NLO.
    Right: Decomposition of the full NLO result in its contributions defined in Eqs.~(\ref{eq:xsec_contribs}).
    \label{fig:pT2}
  }
\end{figure*}

\begin{figure*}[t!]
  \centering
  \includegraphics[width=.45\linewidth]{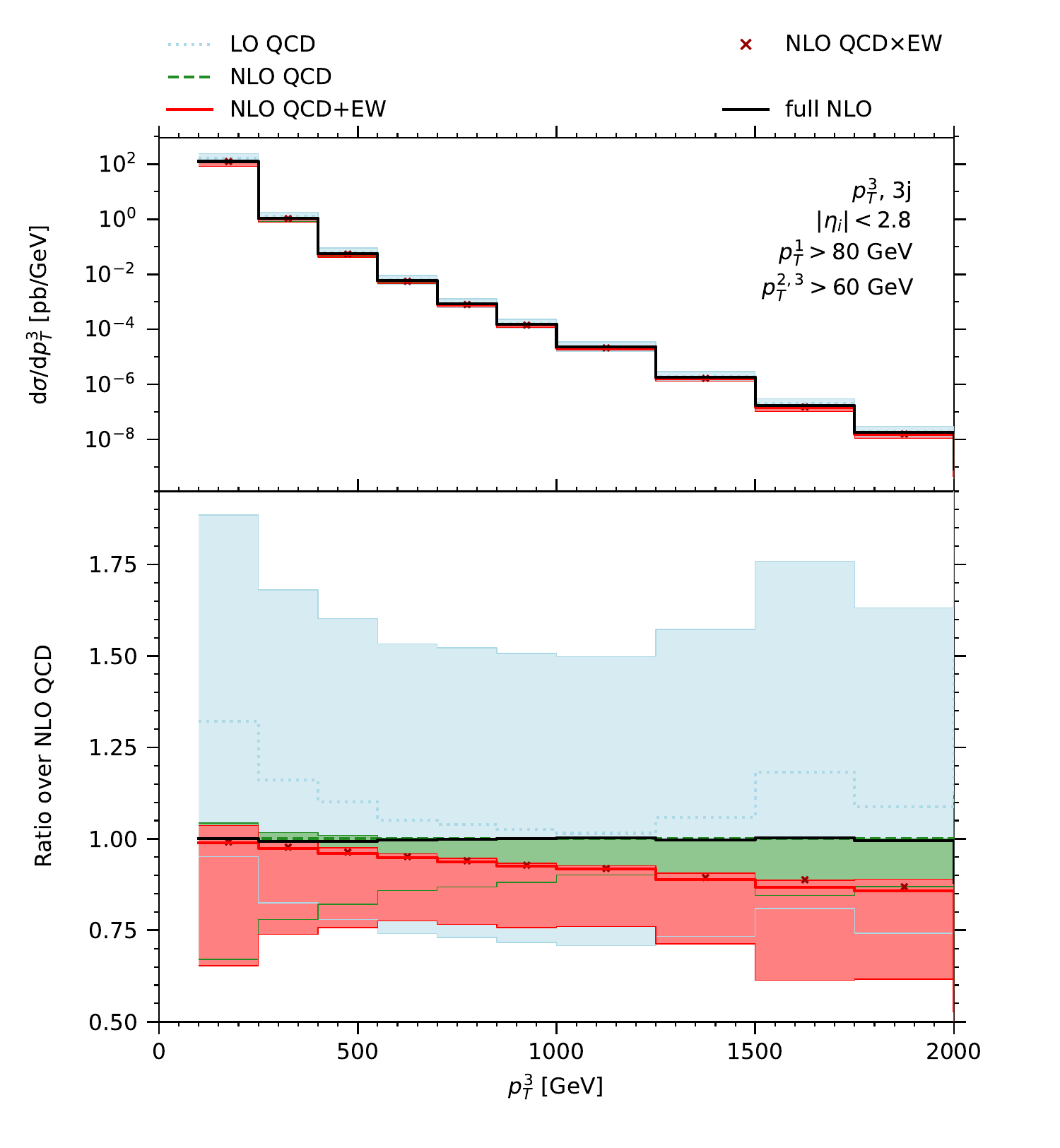}
  \includegraphics[width=.45\linewidth]{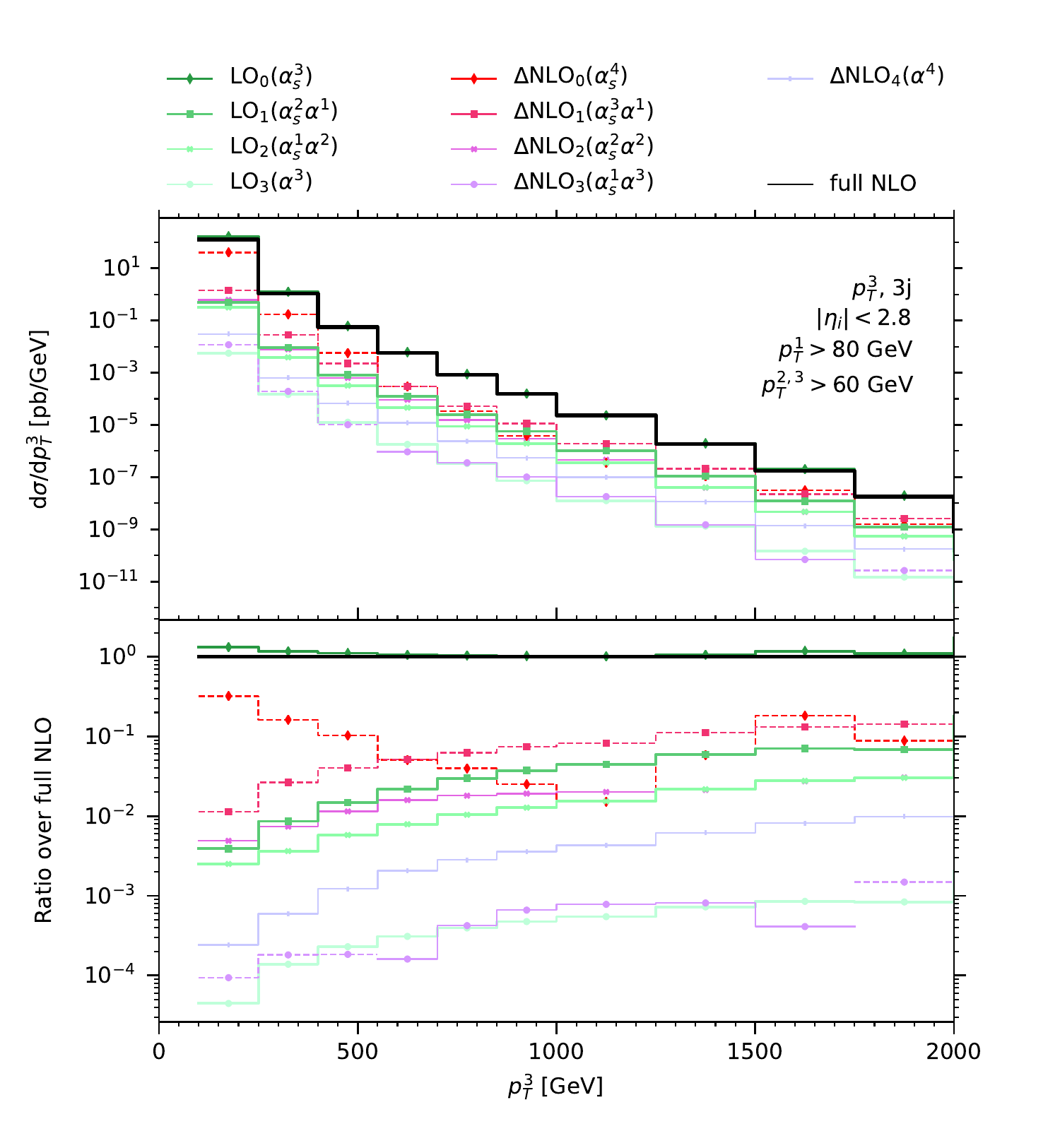}
  \caption{
    Third jet transverse momentum in three-jet production.
    Left: Theoretical uncertainties at LO, NLO QCD, NLO QCD+EW and full NLO.
    Right: Decomposition of the full NLO result in its contributions defined in Eqs.~(\ref{eq:xsec_contribs}).
    \label{fig:pT3}
  }
\end{figure*}

Figures \ref{fig:pT1}--\ref{fig:pT3} show the three-jet 
cross section differential in the transverse momentum of the
leading, subleading and third hardest jet, respectively.
The left hand side panel details the scale uncertainties and 
relative magnitudes of the LO QCD, the NLO \QCDpEW, the 
NLO \QCDtEW\ and the complete NLO (full NLO) result in
comparison to the NLO QCD prediction. 
Similarly, the right hand side panel details the relative contributions 
from the various LO and NLO contributions to the full NLO result
for the central scale choice. Note, while positive sub-contributions
are represented by a solid line, negative parts are indicated
by a dashed line and their corresponding absolute value is displayed
here. 

In all three distributions we confirm the substantial shape correction 
and improvement on the scale uncertainty through the NLO QCD corrections 
observed in earlier calculations of these quantities \cite{Nagy:2003tz}. 
The NLO EW corrections themselves lead to the well-known negative 
corrections of EW Sudakov-type in the high-transverse momentum regime, 
reaching $-10\%$ for the leading, $-15\%$ for the second and $-15\%$ for the
third hardest jet at $\pT=2\,\text{TeV}$. The very good agreement of the
additive and multiplicative combination of QCD and electroweak corrections
indicates a negligible size of the relative $\order(\alphaS\alpha)$ corrections.
The remaining subleading LO and NLO contributions, however, cancel 
the effect of the next-to-leading order electroweak corrections 
almost completely. 
In fact, at $\pT>2.5\,\text{TeV}$ they grow larger and increase 
the full NLO result beyond the NLO QCD one. 
The driving ingredients here are the $\order(\alphaS^3\alpha)$ $\DNLOone$
terms, the tree-level interference $\order(\alphaS^2\alpha)$ (\LOone) contributions,
followed by the interference at $\order(\alphaS\alpha^2)$ (\LOtwo) and their 
respective EW and QCD corrections at $\order(\alphaS^2\alpha^2)$ 
(\DNLOtwo). 
All other contributions to the full NLO result remain marginal. 
It has to be stressed that this cancellation is accidental and 
highly observable dependent and cannot be inferred to hold for any 
other observable, or indeed for the same observable in a different 
fiducial phase space. Lastly we note, that by the inclusion of NLO EW
corrections the uncertainty estimates obtained by QCD scale variations
increases wrt. the NLO QCD result, however, still being significantly
smaller than for the LO QCD prediction. 

\begin{figure*}[t!]
  \centering
  \includegraphics[width=.45\linewidth]{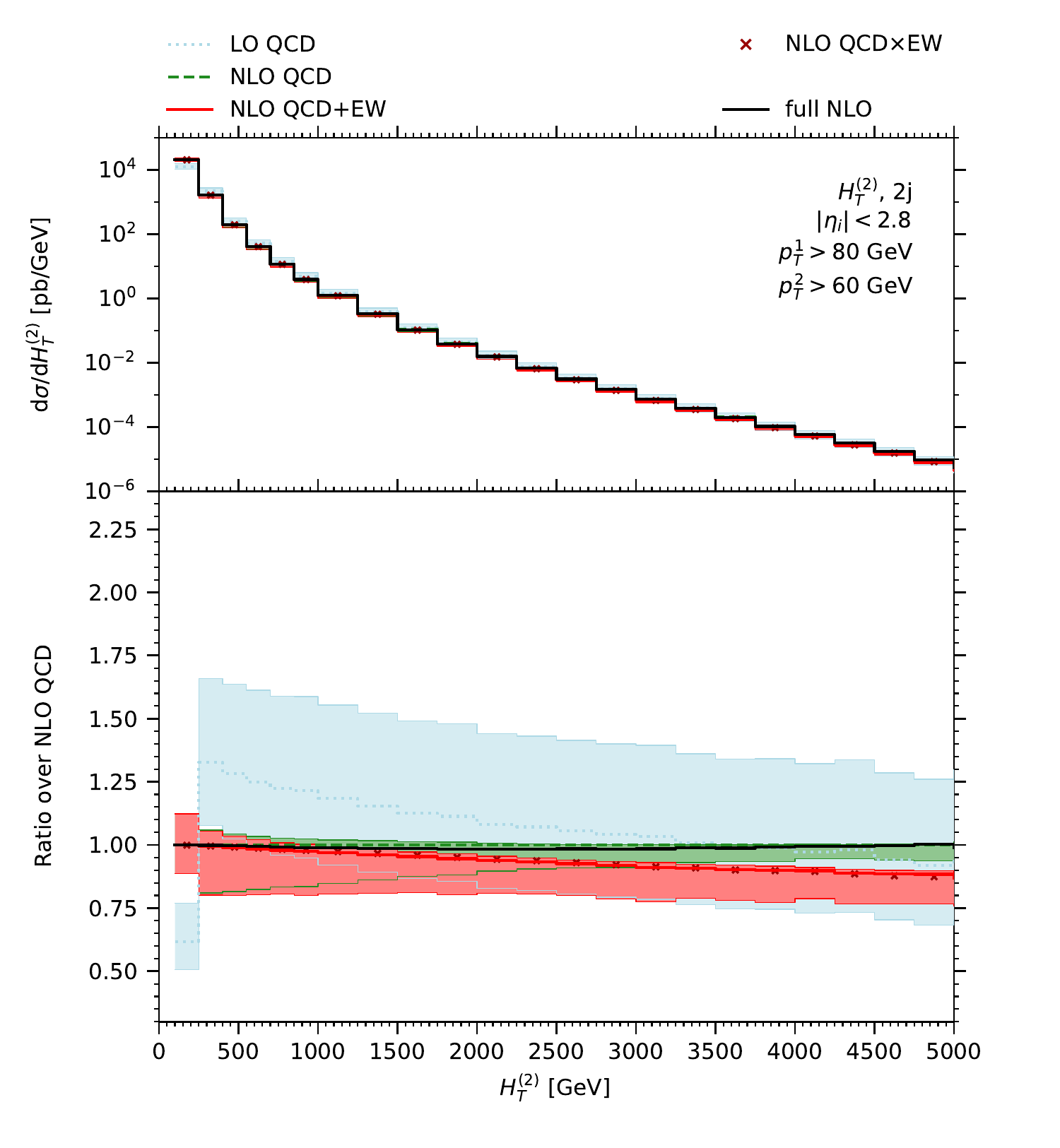}
  \includegraphics[width=.45\linewidth]{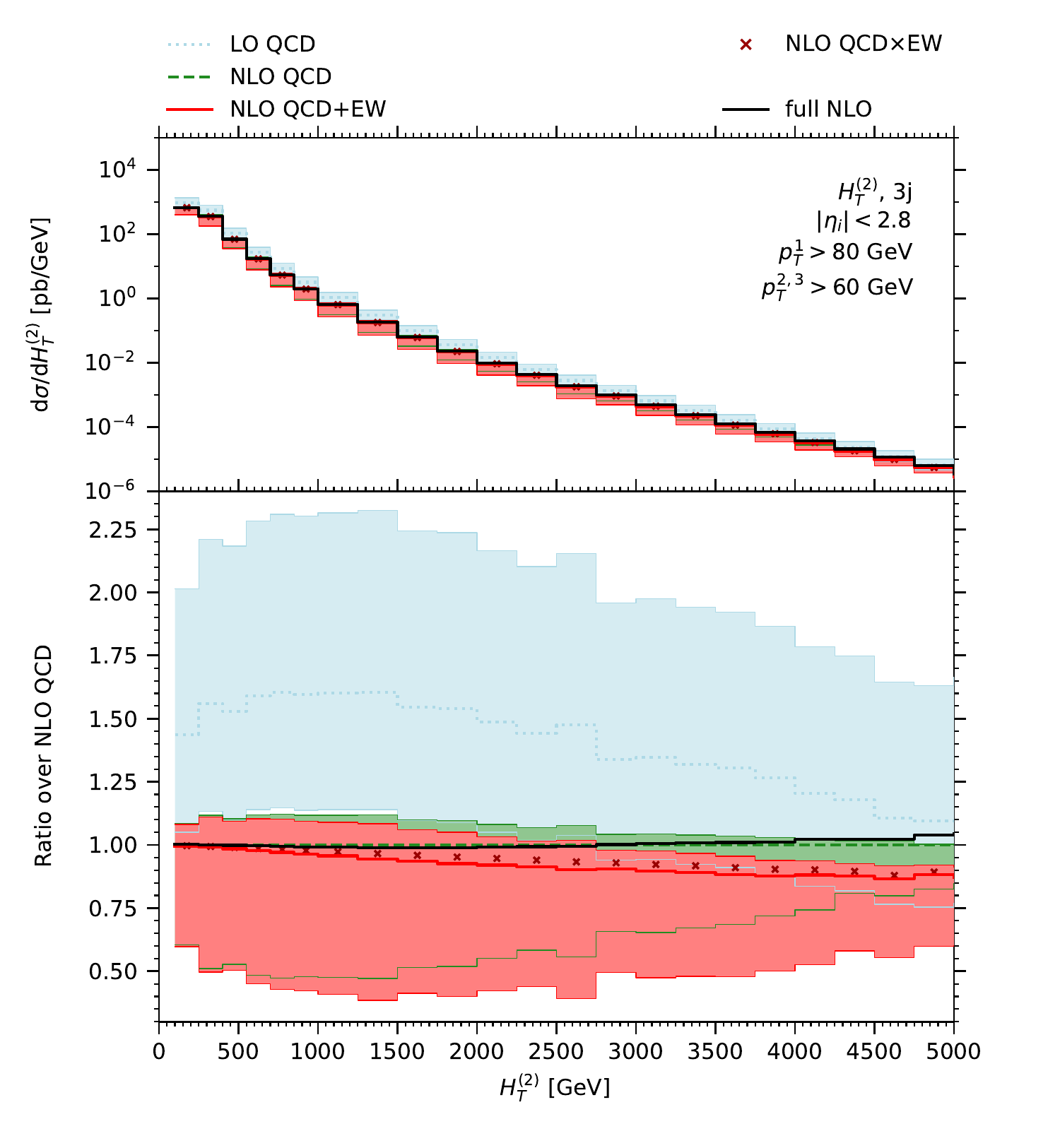}
  \caption{
    The \HTtwo distribution in two- and three-jet production at the LHC shown in the left and right panel,
    respectively. Besides the full NLO prediction the central results and scale uncertainty bands for LO
    and NLO QCD, NLO \QCDpEW\ and NLO  \QCDpEW\ are shown. 
    \label{fig:HT2}
  }
\end{figure*}

Figure \ref{fig:HT2} now displays the results for the scalar sum of the 
leading and subleading jet transverse momenta, \HTtwo, in two- and three-jet
events. While the latter represents a novel result from our full NLO
three-jet calculation, the first is obtained from a dijet computation
with identical parameter settings, scale choices and PDFs. Qualitatively,
the \HTtwo\ distributions exhibit the same features as 
the leading and subleading jet transverse momentum distributions presented
before. 
While the scale uncertainties are shrunk going from LO to NLO QCD, 
the electroweak corrections show the expected Sudakov behaviour. 
The relative electroweak corrections are of nearly the same magnitude 
for both the two- and the three-jet case. 
This can be understood from the fact that with \HTtwo\ in the TeV 
region, where the electroweak corrections become sizeable, the 
additional third jet in the three-jet case is predominantly soft and 
near the jet threshold. In this limit, higher order QCD and EW corrections
should factorise. Further, we note that for both distributions the
additive and multiplicative combination of NLO QCD and EW corrections
give compatible results. As has been observed before in the jet transverse
momenta, including electroweak contributions somewhat increases the uncertainty
wrt.\ NLO QCD. 

\begin{figure*}[t!]
  \centering
  \begin{minipage}{0.03\textwidth}
    \rotatebox{90}{dijet production}
  \end{minipage}
  \hfill
  \begin{minipage}{0.95\textwidth}
    \includegraphics[width=.33\linewidth]{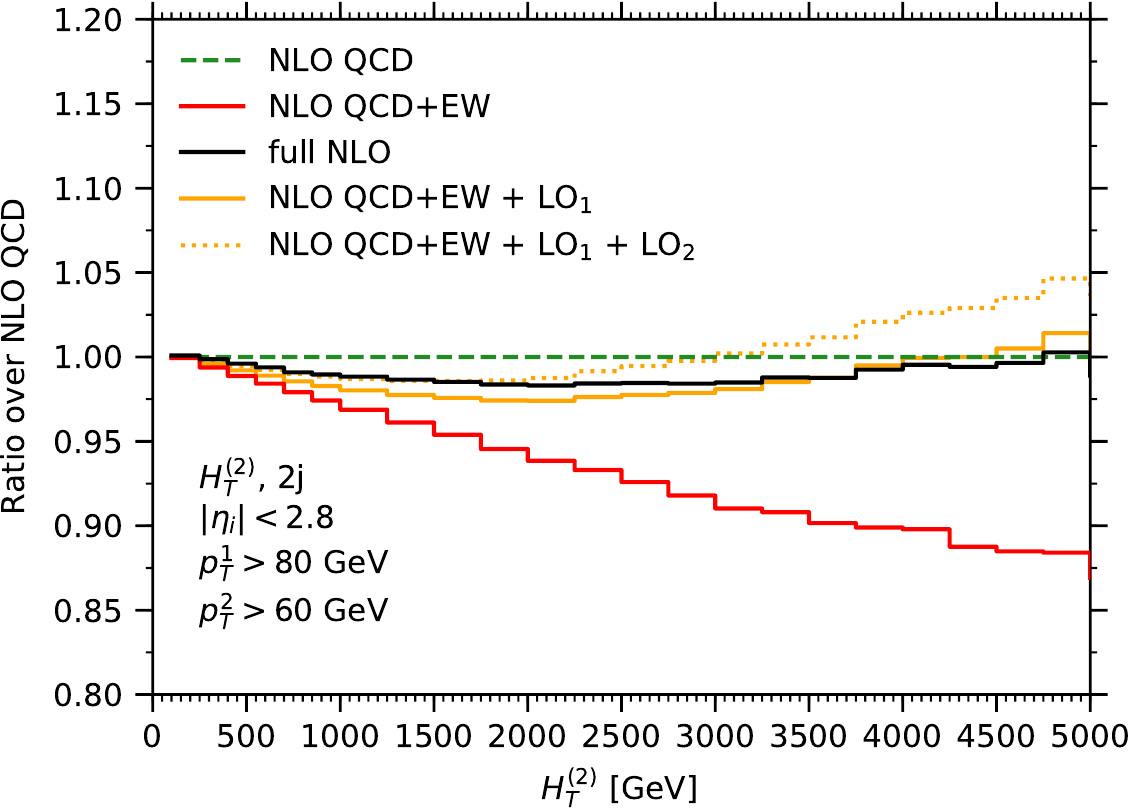}
    \includegraphics[width=.33\linewidth]{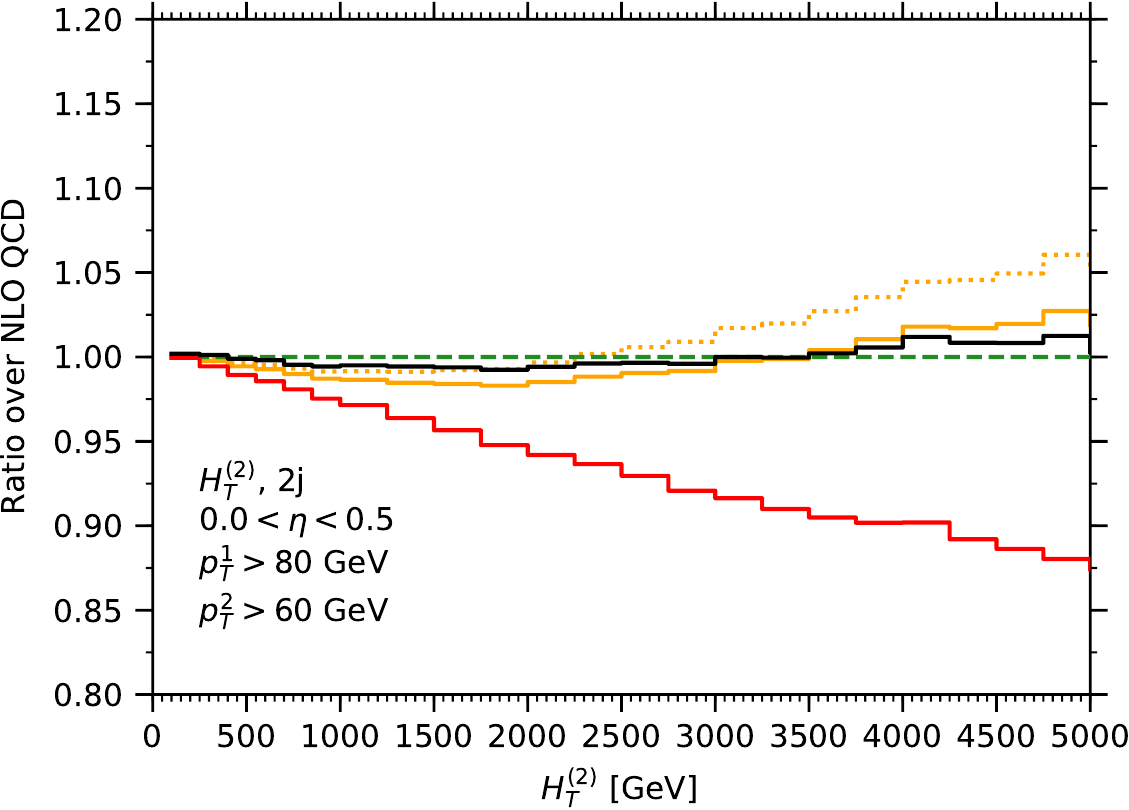}
    \includegraphics[width=.33\linewidth]{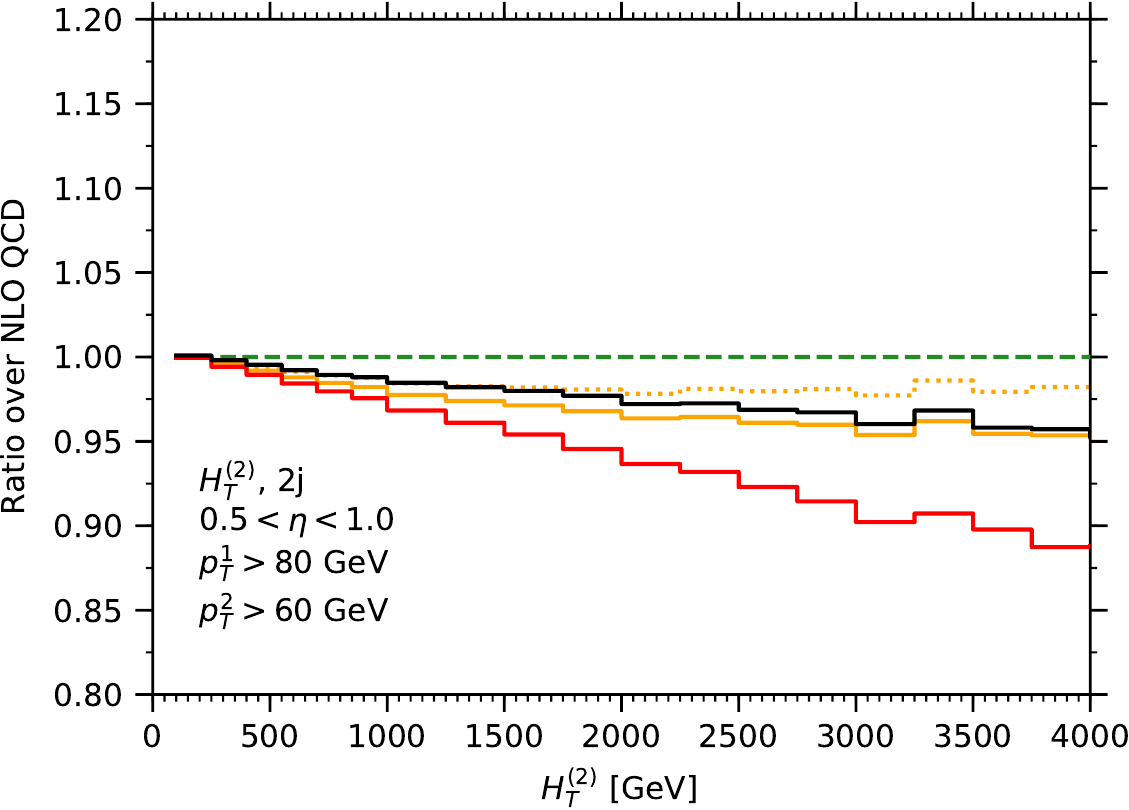}\\
    \includegraphics[width=.33\linewidth]{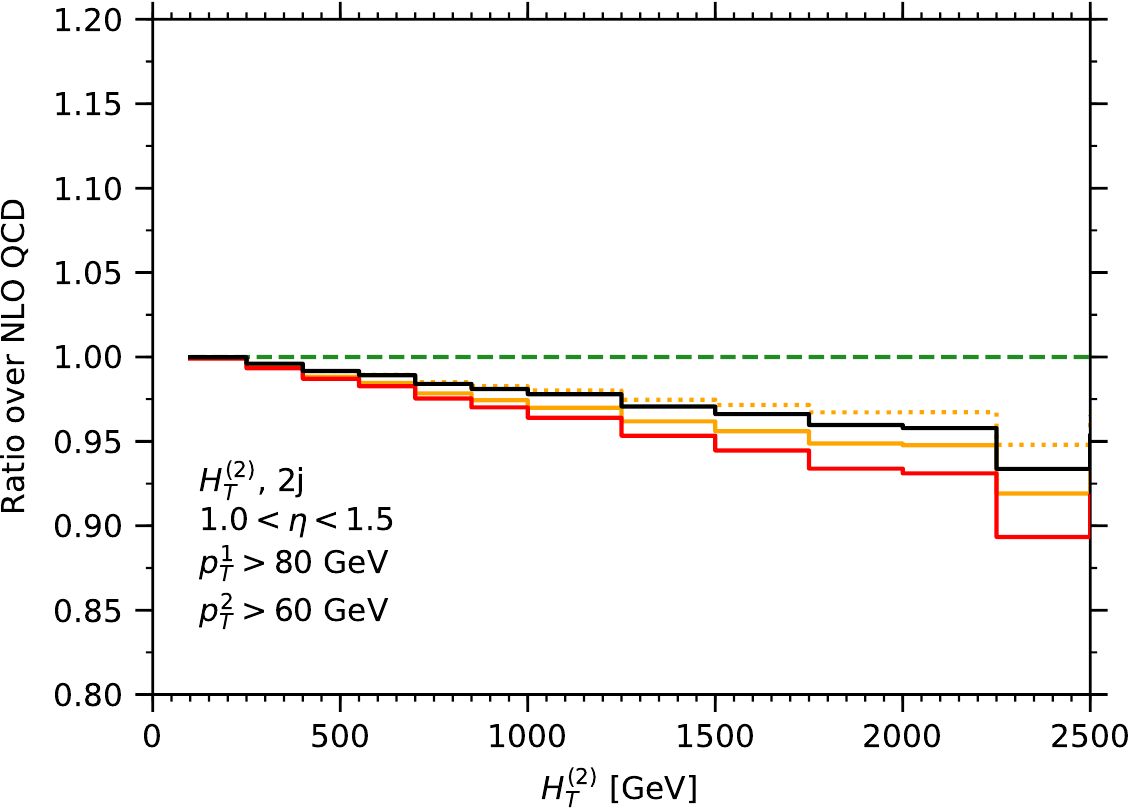}
    \includegraphics[width=.33\linewidth]{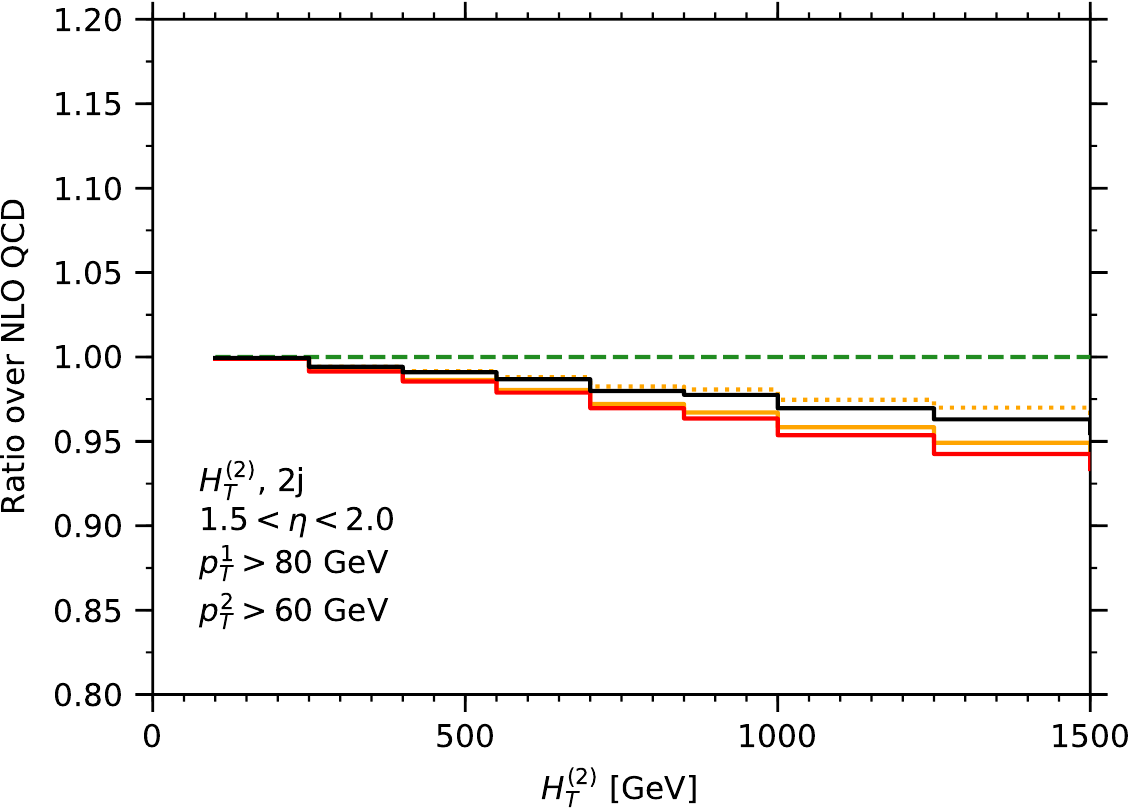}
    \includegraphics[width=.33\linewidth]{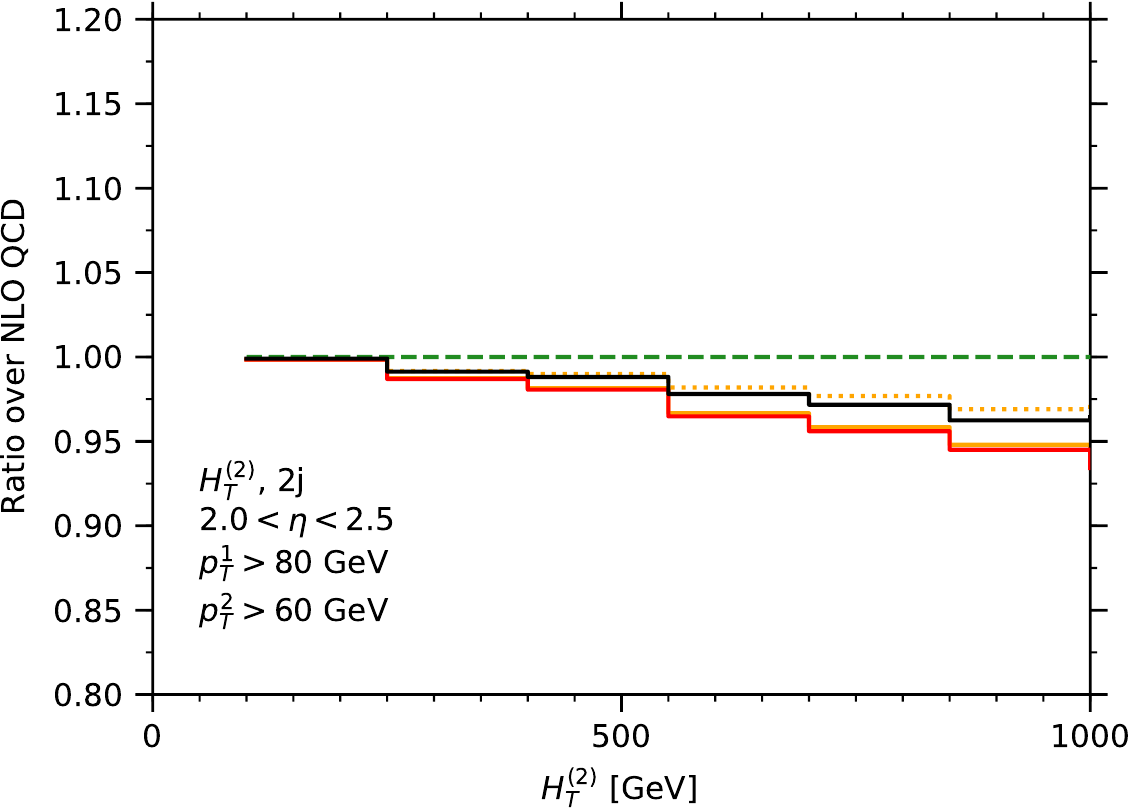}
  \end{minipage}
  \vspace{.15cm}
  \\\hrule
  \vspace{.3cm}
  \begin{minipage}{0.03\textwidth}
    \rotatebox{90}{three jet production}
  \end{minipage}
  \hfill
  \begin{minipage}{0.95\textwidth}
    \includegraphics[width=.33\linewidth]{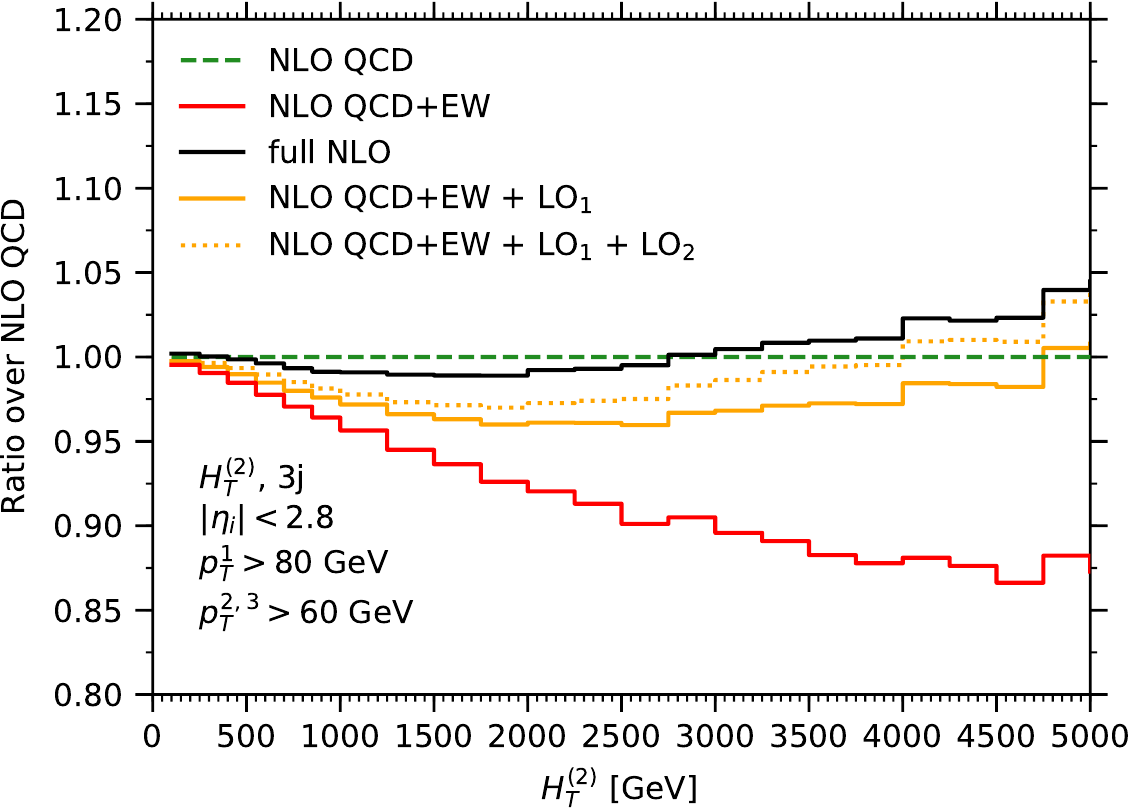}
    \includegraphics[width=.33\linewidth]{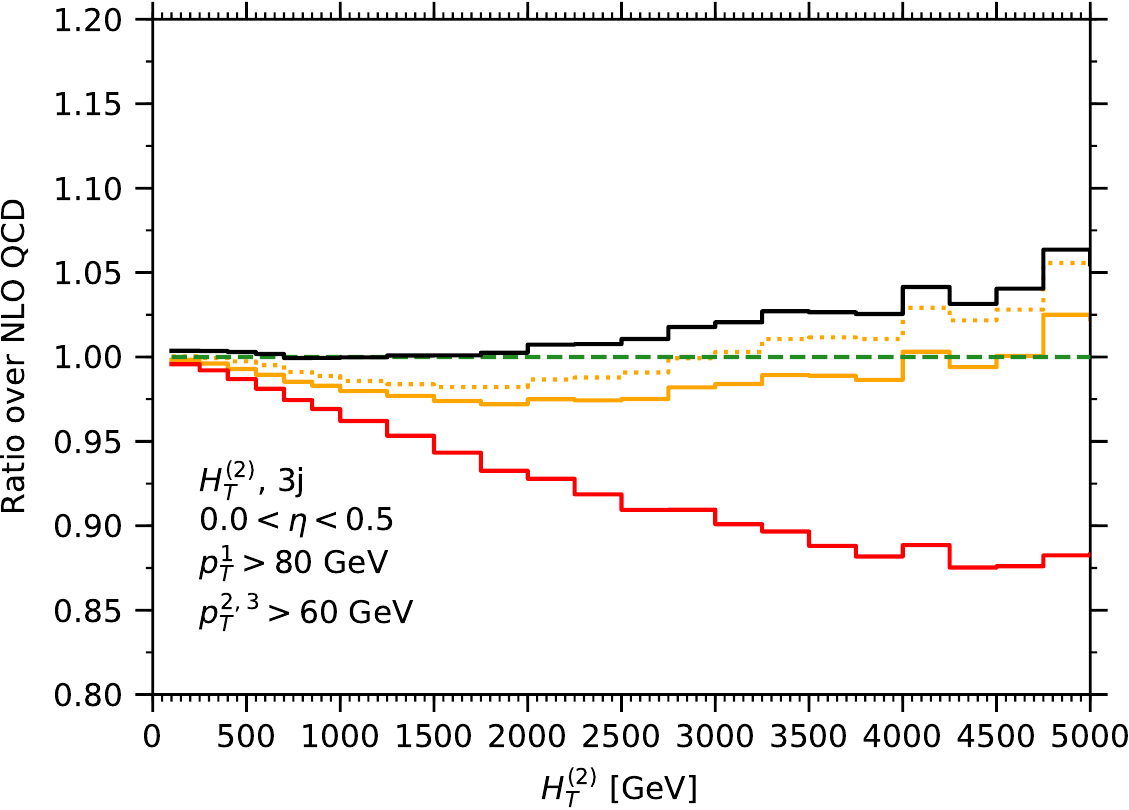}
    \includegraphics[width=.33\linewidth]{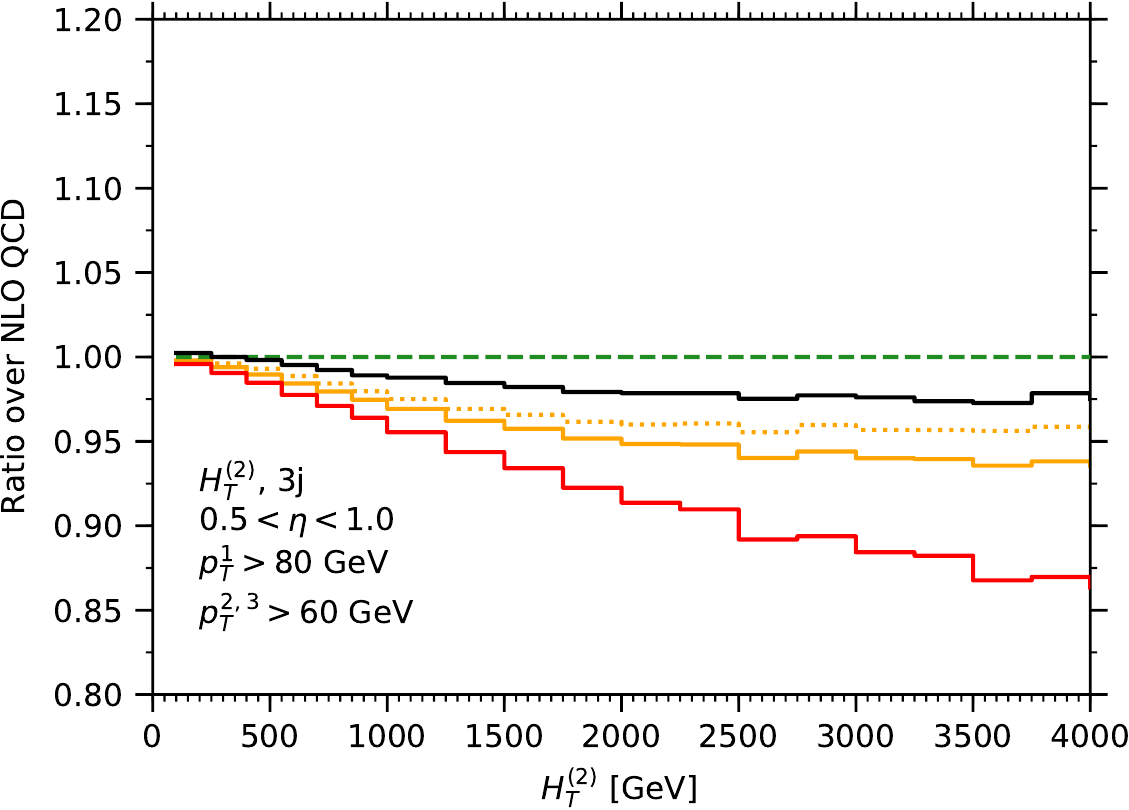}\\
    \includegraphics[width=.33\linewidth]{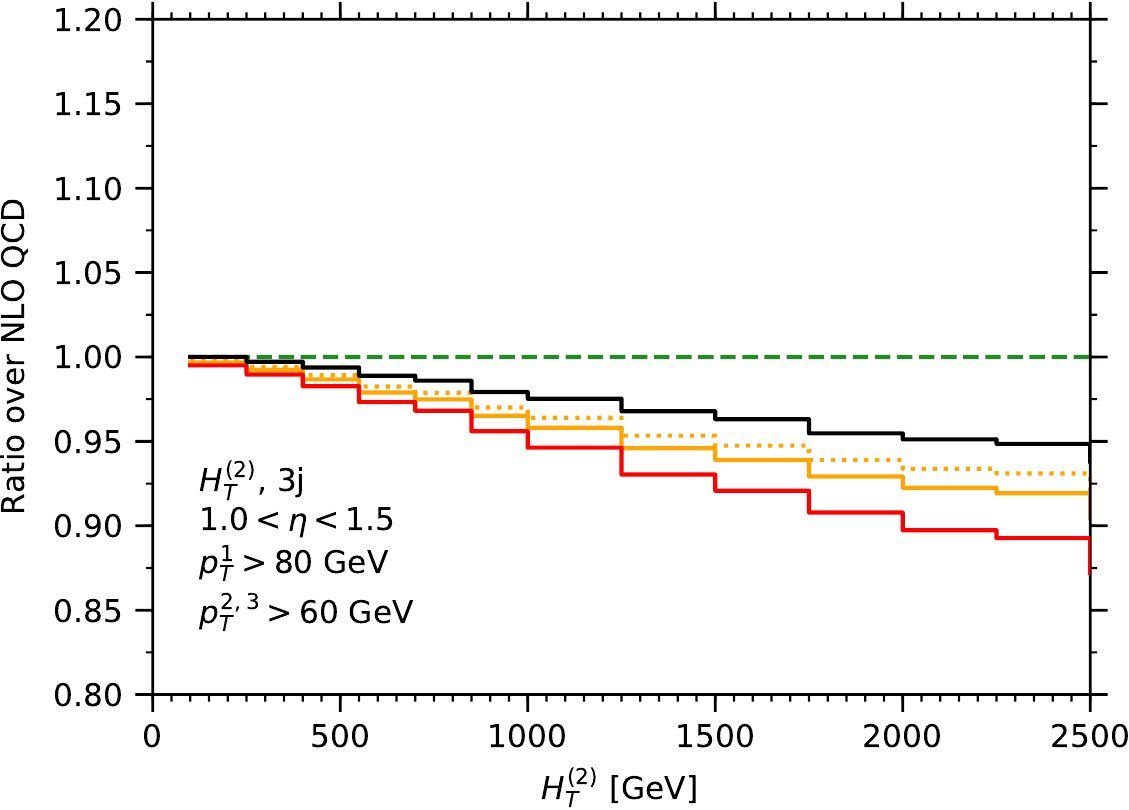}
    \includegraphics[width=.33\linewidth]{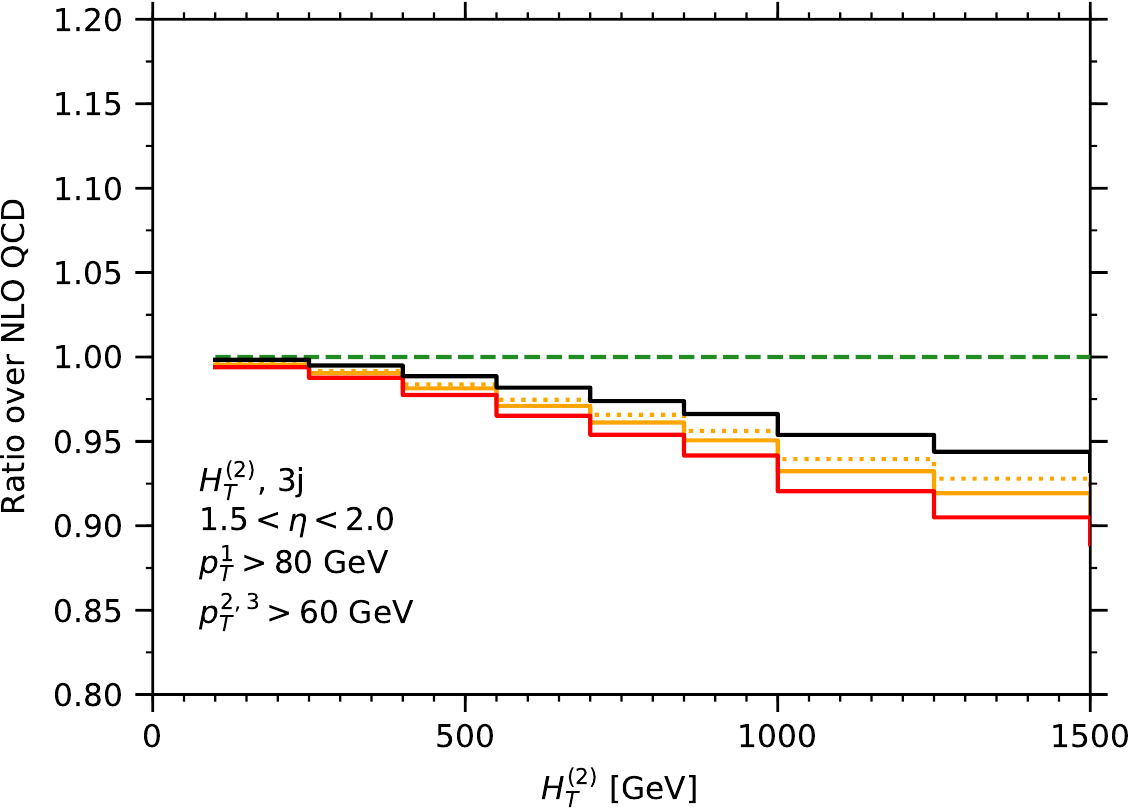}
    \includegraphics[width=.33\linewidth]{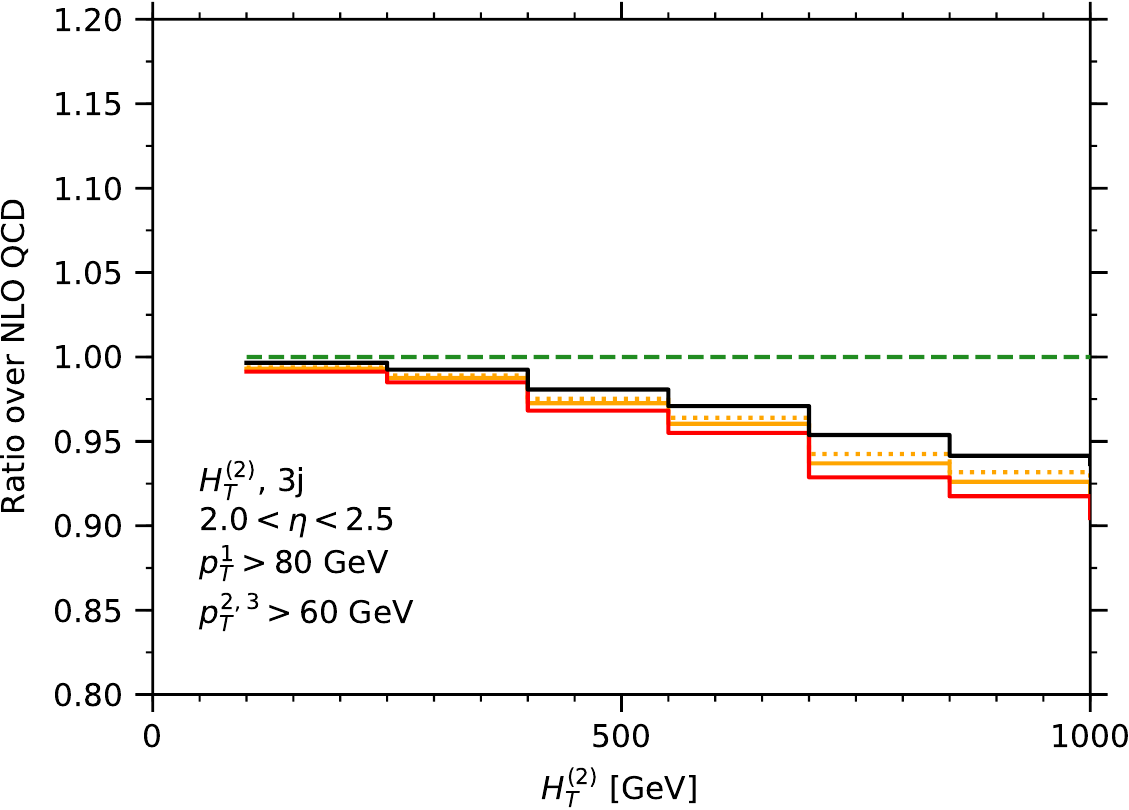}
  \end{minipage}
  \caption{
    Scalar sum of leading and subleading jet transverse momentum, \HTtwo, in two 
    (top panel) and three (bottom panel) 
    jet production, decomposed into contributions from several ranges
    of $\eta=|\eta_1-\eta_2|/2$.
    Shown are the NLO QCD, NLO \QCDpEW\ and full NLO result as well as the
    subleading Born contributions \LOone\ and \LOtwo.
    \label{fig:HT2_doublediff}
  }
\end{figure*}

Upon inclusion of the additional subleading LO and NLO contributions
NLO EW effects get cancelled and the full NLO result gets
very close to the NLO QCD prediction. Interestingly, this is true both
for the two- and three-jet case. However, this cancellation is accidental 
and highly dependent on the observable and the phase space considered. 
To illustrate this observation, Figure \ref{fig:HT2_doublediff} 
shows the same observable, \HTtwo, in different regions of absolute 
pseudorapidity of the leading two-jet system, i.e. $\eta=|\eta_1-\eta_2|/2$.
In the central region, which dominates the inclusive result, 
the subleading contributions, dominated by \LOone\ in 
both the two- and three-jet case, have a large positive effect on the 
cross section.
The more forward \HTtwo\ is considered, however, the smaller especially 
the \LOone\ terms become and the closer the full NLO result is to the 
NLO \QCDpEW\ one. 
This was already observed in \cite{Dittmaier:2012kx}. 
In this region, also qualitative differences between the two- and three-jet case are apparent. 
While the further subleading contributions are negative wrt.\ the 
NLO \QCDpEW\ result in the dijet case, they are positive wrt.\ the 
NLO \QCDpEW\ result in the three-jet case.

\begin{figure}[t!]
  \centering
  \includegraphics[width=0.9\linewidth]{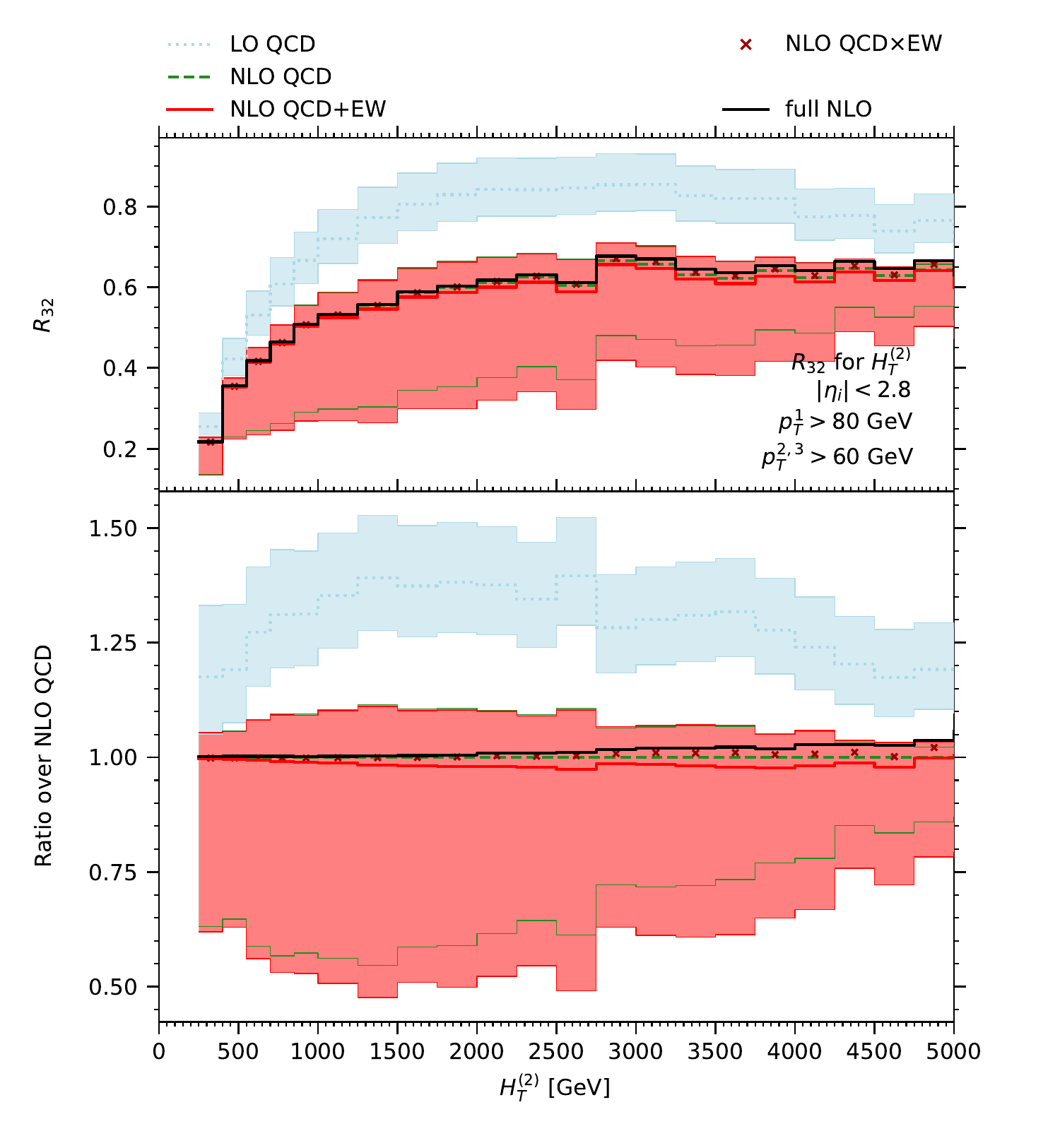} 
  \caption{
    The \Rtt\ observable differential in  \HTtwo. Upper panel: Predictions
    at LO and NLO QCD, NLO \QCDpEW, NLO \QCDtEW\ and full NLO in the Standard
    Model. Lower panel: Related relative corrections wrt.\ the central NLO QCD result.
    \label{fig:R32HT2}
  }
\end{figure}

With full NLO calculations for two- and three-jet production at hand we
turn to the central observable of this letter, the three-jet-over-two-jet ratio, \Rtt. 
This particular observable has attracted interest, as large parts
of the experimental and theoretical uncertainties in the inclusive three- and two-jet
cross sections cancel in the ratio, allowing for a competitive measurement of
the strong coupling $\alphaS$~\cite{Aad:2011tqa,Chatrchyan:2013txa}.
Here we consider \Rtt\ differential in \HTtwo, the scalar sum of the leading-
and subleading-jet transverse momenta presented above, i.e.\ 
\begin{equation}
  \begin{split}
    \Rtt(\HTtwo) 
    \,=\,\frac{\done\sigma_{3j}/\done\HTtwo}
              {\done\sigma_{2j}/\done\HTtwo}\;.
  \end{split}
\end{equation}
The scale uncertainties are computed by synchronous variations of
numerator and denominator. Our results are presented in Figure~\ref{fig:R32HT2}. 

\begin{figure*}[t!]
  \centering
  \includegraphics[width=.32\linewidth]{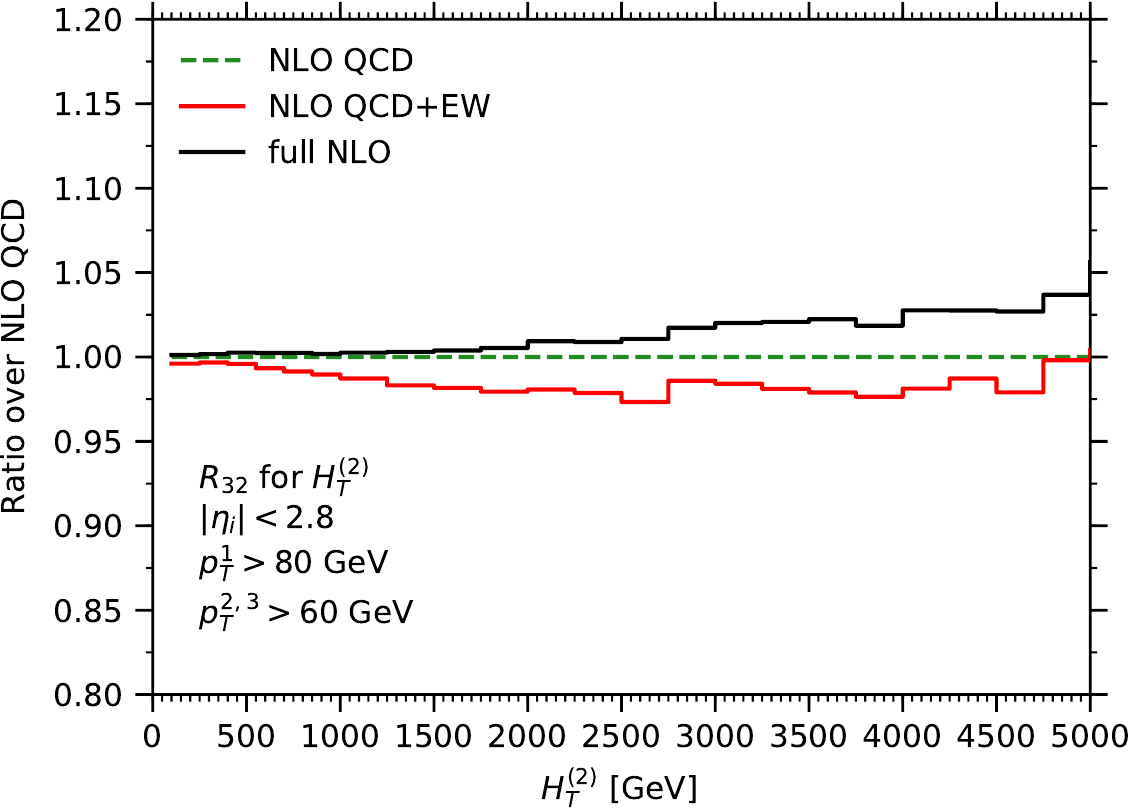}
  \includegraphics[width=.32\linewidth]{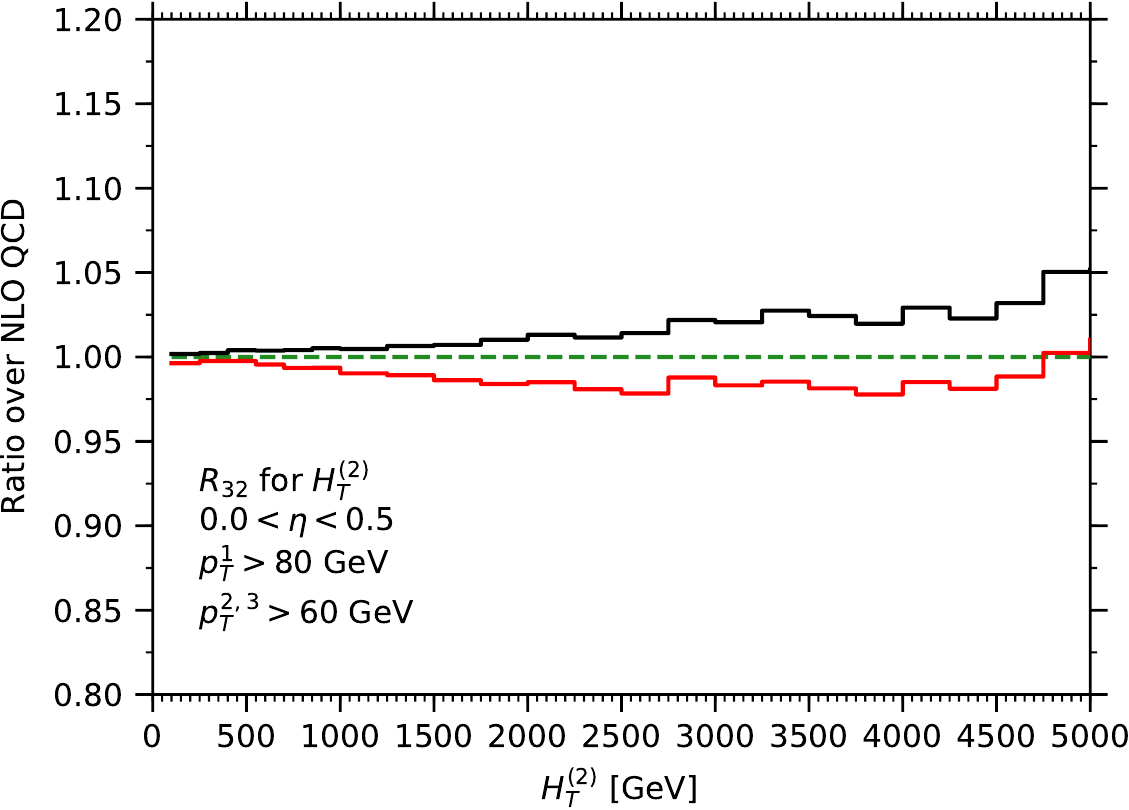}
  \includegraphics[width=.32\linewidth]{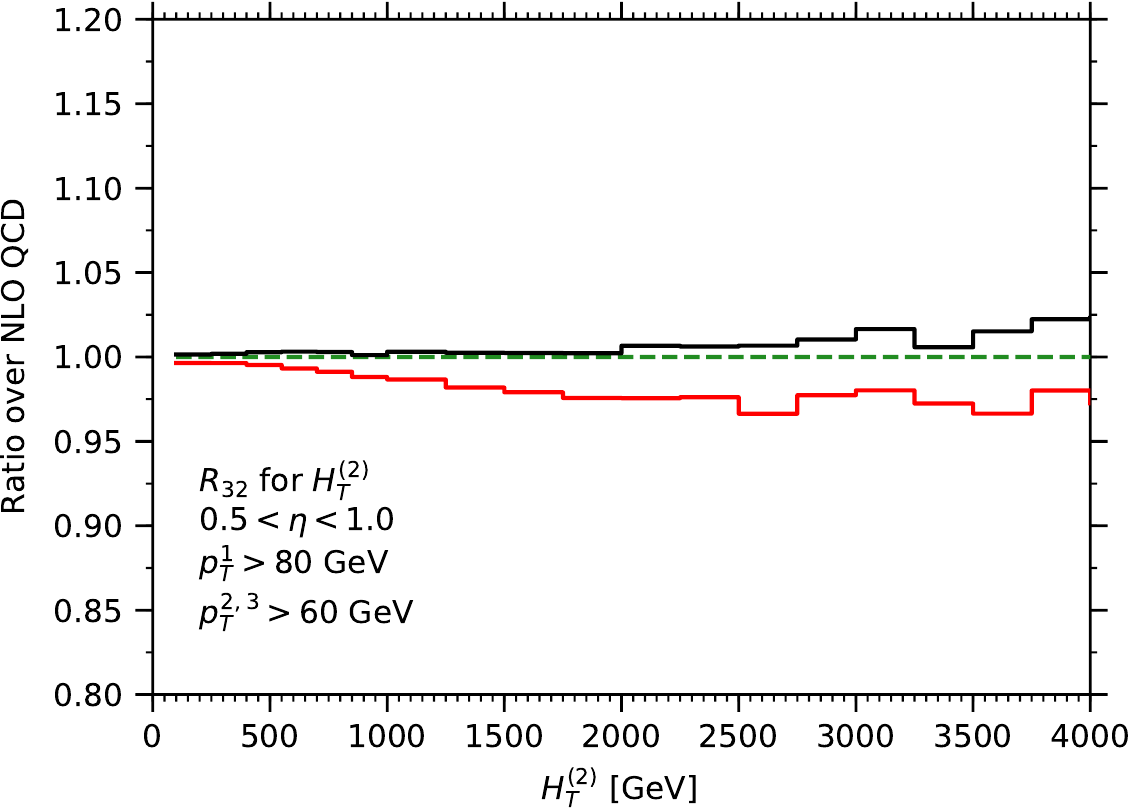}
  \includegraphics[width=.32\linewidth]{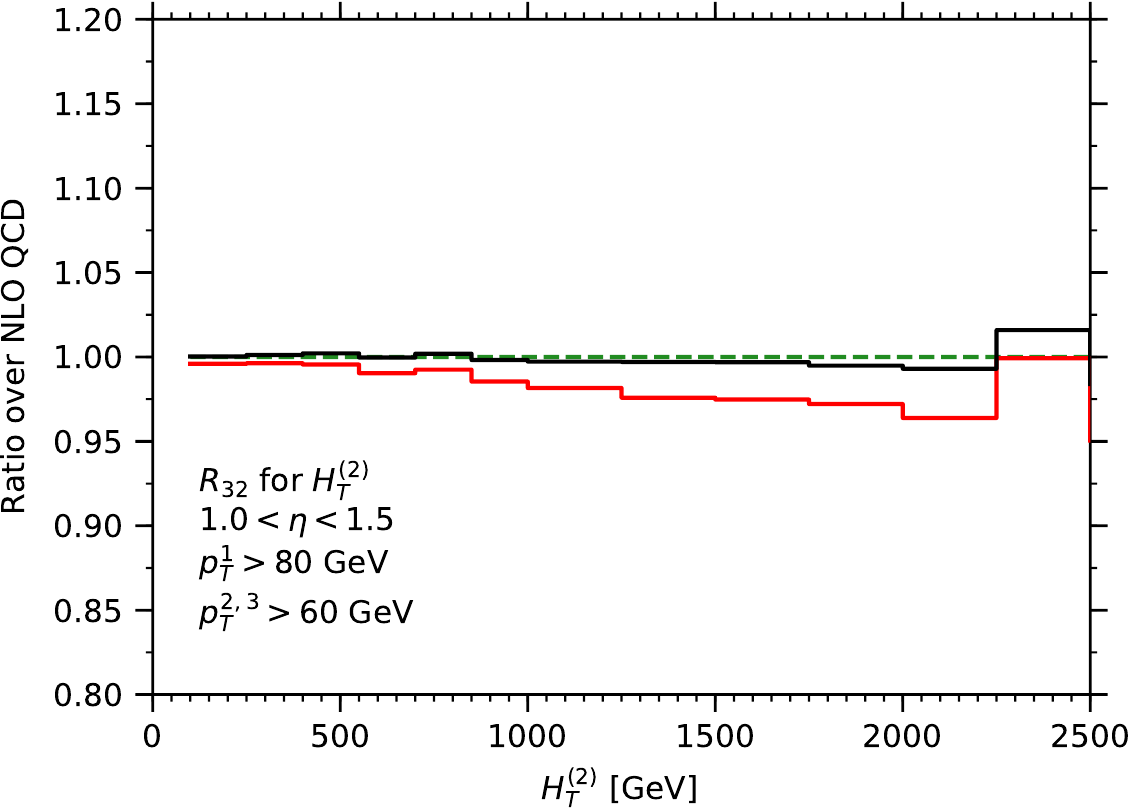}
  \includegraphics[width=.32\linewidth]{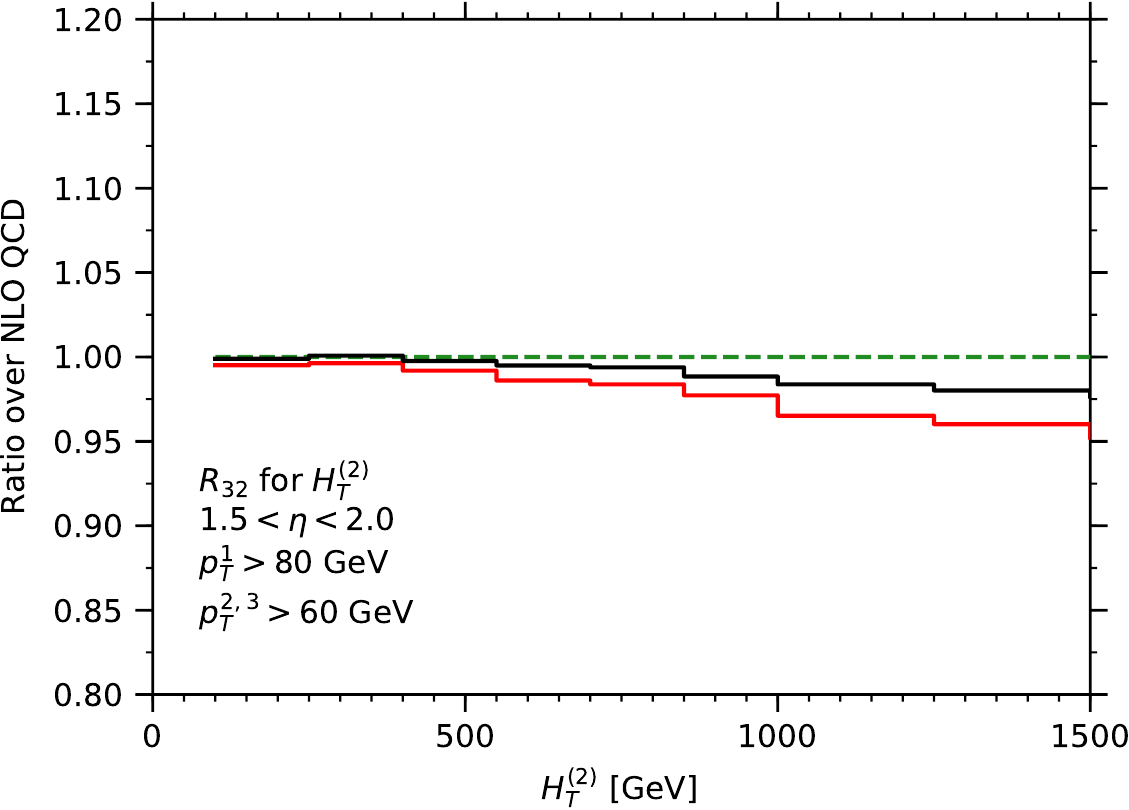}
  \includegraphics[width=.32\linewidth]{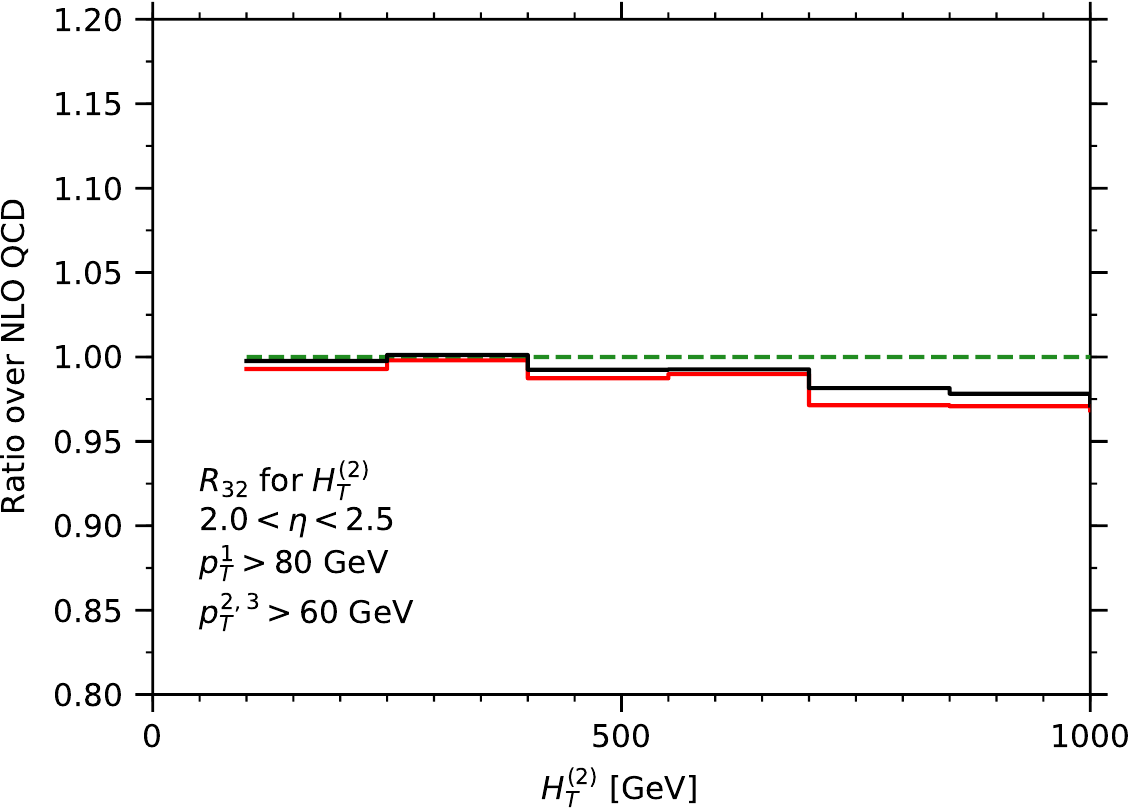}
  \caption{
    The \Rtt\ observable at NLO \QCDpEW\ and full NLO differential in  \HTtwo for different pseudo-rapidity selections
    and in comparison to the corresponding NLO QCD result.
    \label{fig:R32HT2_doublediff}
  }
\end{figure*}

We find that as the individual input distributions receive only minute EW
corrections, resulting in the NLO QCD predictions to agree with the full NLO,
also their ratio is very stable. However, as emphasised before, accidental
cancellations of individually much larger terms is in action for this observable. 
Therefore, we present in Figure \ref{fig:R32HT2_doublediff} results differential
in various pseudorapidity regions, with $\eta =|\eta_1-\eta_2|/2$. 
As before, the inclusive result is dominated by the most central 
pseudorapidity slices, and they exhibit the same characteristics. 
In the slightly more forward regions, between $0.5\leq\eta\leq 2$, the input 
distributions of Figure \ref{fig:HT2_doublediff} do not exhibit this 
almost complete cancellation of corrections any longer. 

For the cross-section ratio \Rtt\ the net effect is nonetheless the same and the
residual corrections of electroweak and subleading origin are very small. 
Their contributions largely factorise in the numerator and denominator
and, thus, cancel in the ratio. 
Hence, the full NLO result is in very good agreement with the NLO QCD prediction 
for this observable. 
This very much confirms the particular usefulness of \Rtt\ for the determination
of the strong coupling. 

\section{Conclusions}
\label{sec:conclusions}

In this letter we have presented the evaluation of the full set of Standard Model NLO corrections
to three-jet production at the LHC. Besides the dominating QCD corrections of $\order(\alphaS^4)$
this comprises all (mixed) electroweak tree-level contributions up to $\order(\alpha^3)$ as well as
all (mixed) one-loop and real-corrections up to $\order(\alpha^4)$. As jet constituents we consider
besides quarks and gluons also photons and charged leptons. However, for the considered event
selections contributions from final states containing leptons are practically irrelevant. All calculations
have been performed in an automated manner within the \Sherpa\ event generation framework, with \Recola\
providing the renormalised virtual corrections. 

For the jet transverse momentum distributions and the related $\HTtwo$ variable we observe a
compensation of the electroweak Sudakov-type suppression of high-$\pT$ events when including
subleading electroweak tree-level and one-loop contributions. In fact, for leading jet
transverse momenta above $2.5\,\TeV$ a resulting positive correction of $10-15\%$ wrt. the NLO
QCD prediction is observed. However, the mentioned compensation is specific for the fiducial
phase-space region considered. In particular for jet production away from central rapidity
we observe sizeable effects upon inclusion of the full set of (mixed) electroweak corrections.
This nicely illustrates the importance of considering the complete set of NLO Standard Model
corrections in predictions for the three-jet production process at the LHC. 

As a first application of our calculation we have considered the ratio of three- over two-jet
production differential in $\HTtwo$. This variable proves to be very stable against electroweak
corrections, confirming its particular usefulness in the determination of the strong coupling
constant $\alphaS$.

\begin{acknowledgements}
  This work has received funding from the European Union's Horizon 2020 research and innovation programme as part of the Marie Sk\l{}odowska-Curie Innovative Training Network MCnetITN3 (grant agreement no. 722104). SS acknowledges support through the Fulbright-Cottrell Award and from BMBF (contracts 05H15MGCAA and 05H18MGCA1). 
  MS acknowledges the support of the Royal Society through the award 
  of a University Research Fellowship.
  MR is supported by the Research Training Group GRK 2044 of the German Research Foundation (DFG).
\end{acknowledgements}

\bibliographystyle{spphys}
\bibliography{refs.bib}

\end{document}